\documentclass{elsarticle}
\usepackage{color}
\usepackage[pagewise]{lineno}
\usepackage{amsmath}
\usepackage{units}
\usepackage{xspace}
\usepackage{subfig}
\usepackage{graphicx}
\usepackage{units}
\usepackage{url}
\usepackage{amssymb}
\usepackage{longtable}
\usepackage{epsfig}
\usepackage{units}
\usepackage{lineno}
\usepackage{color}
\usepackage{booktabs}
\usepackage{color}

\newcommand*\patchAmsMathEnvironmentForLineno[1]{%
  \expandafter\let\csname old#1\expandafter\endcsname\csname #1\endcsname
  \expandafter\let\csname oldend#1\expandafter\endcsname\csname end#1\endcsname
  \renewenvironment{#1}%
     {\linenomath\csname old#1\endcsname}%
     {\csname oldend#1\endcsname\endlinenomath}}%
\def\EeV{\ifmmode {\mathrm{\ Ee\kern -0.1em V}}\else
                   \textrm{Ee\kern -0.1em V}\fi}%
\def\eV{\ifmmode {\mathrm{\ e\kern -0.1em V}}\else
                   \textrm{e\kern -0.1em V}\fi}%
\def\gcm{\ifmmode {\mathrm{g/cm}^2}\else
                   {g/cm$^2$}\fi\xspace}%
\def\Xmax{\ifmmode {X_\mathrm{max}}\else
                   {$X_\mathrm{max}$}\fi\xspace}%

\begin{document}
\begin{frontmatter}
\title{ {\bf The Lateral Trigger Probability function for the Ultra-High Energy Cosmic Ray Showers detected by the Pierre Auger Observatory}}

\author[]{\small 
\par\noindent
{\bf The Pierre Auger Collaboration} \\
P.~Abreu$^{74}$, 
M.~Aglietta$^{57}$, 
E.J.~Ahn$^{93}$, 
I.F.M.~Albuquerque$^{19}$, 
D.~Allard$^{33}$, 
I.~Allekotte$^{1}$, 
J.~Allen$^{96}$, 
P.~Allison$^{98}$, 
J.~Alvarez Castillo$^{67}$, 
J.~Alvarez-Mu\~{n}iz$^{84}$
% ...

\par\noindent
$^{1}$ Centro At\'{o}mico Bariloche and Instituto Balseiro (CNEA-
UNCuyo-CONICET), San Carlos de Bariloche, Argentina \\
% ...
$^{19}$ Universidade de S\~{a}o Paulo, Instituto de F\'{\i}sica, S\~{a}o 
Paulo, SP, Brazil \\
% ...
$^{33}$ Laboratoire AstroParticule et Cosmologie (APC), 
Universit\'{e} Paris 7, CNRS-IN2P3, Paris, France \\
% ...
$^{57}$ Istituto di Fisica dello Spazio Interplanetario (INAF),
 Universit\`{a} di Torino and Sezione INFN, Torino, Italy \\
% ...
$^{67}$ Universidad Nacional Autonoma de Mexico, Mexico, D.F., 
Mexico \\
% ...
$^{74}$ LIP and Instituto Superior T\'{e}cnico, Lisboa, Portugal \\
% ...
$^{84}$ Universidad de Santiago de Compostela, Spain \\
% ...
$^{93}$ Fermilab, Batavia, IL, USA \\
% ...
$^{96}$ New York University, New York, NY, USA \\
% ...
$^{98}$ Ohio State University, Columbus, OH, USA \\
% ...
\par\noindent
($\dagger$) Deceased \\
(a) at Konan University, Kobe, Japan \\
}

\begin{abstract}

In this paper we introduce the concept of Lateral Trigger
Probability (LTP) function, i.e., the probability for an extensive air shower (EAS)
to trigger an individual detector of a ground based array as a function of distance to the shower axis, taking into account energy, mass and direction of the primary cosmic ray. We apply this concept to the surface array of the Pierre Auger Observatory 
 consisting of a 1.5 km spaced grid of about 1600 water Cherenkov stations. 
   Using Monte Carlo simulations of ultra-high energy showers
 the LTP functions are derived for energies in the range between $10^{17}$ and $10^{19}$~eV and zenith angles up to 65$^\circ$.  
A parametrization combining a step function with an exponential  
is found to reproduce them very well in the considered range of  
energies and zenith angles. 
The LTP functions can also be obtained from data using 
events simultaneously observed by the fluorescence and the surface detector of the Pierre 
Auger Observatory (hybrid events).
We validate the
Monte-Carlo results showing how LTP functions  from data are in good agreement with simulations.

\end{abstract}

\end{frontmatter}

\section{Introduction}
\label{sec:intro}

The Pierre Auger Observatory has been conceived to study the origin and the nature of ultra high-energy cosmic rays.  
Because of the scarcity of the flux at the highest energies, 
their direct measurement from space is technically unfeasible and 
the use  of very large detectors is required at the ground. What can be
observed is the extensive air shower of secondary particles produced in the propagation through the atmosphere.
The Pierre Auger Observatory
is located near Malarg\"ue, Argentina, at 1400 m a.s.l. and it employes  
two independent and complementary measurement techniques~\cite{engineering}. 
The surface array (SD), consisting of about 1600 water Cherenkov detectors 
 on a triangular grid of 1.5 km spacing covering an area of approximately 
3000 km$^2$, records the secondary particles at the ground and thus samples their lateral density distribution. 
The fluorescence detector (FD), consisting of 24 telescopes at four 
sites, overlooks the surface array and observes the longitudinal profile of air showers 
by collecting the   
fluorescence light emitted along the path through the atmosphere~\cite{FD_paper}.  
Unlike the surface detector array with its nearly 100\% duty cycle,  
the FD can only operate on clear and moonless nights 
giving an overall duty cycle of about 13\%~\cite{hybrid_aperture}. As a consequence, only a fraction of showers are observed by both detectors. 
For these events, called hereafter ``hybrid'',   
 the combination of information 
from the surface array and the fluorescence telescopes enhances the
reconstruction capability. Energy and direction reconstruction accuracy of hybrid events is in fact better than 
the one the SD and FD could achieve independently.

One of the main goals of the Pierre Auger Observatory is to measure the flux of cosmic rays at the highest energies.  
This task relies on an accurate determination of the detector exposure for SD-only~\cite{SD_acceptance} and 
hybrid~\cite{hybrid_aperture} operation modes. 
The hybrid exposure is calculated using the simultaneous simulation of FD and SD response.  
Besides the dependence on energy and distance to an FD-site, 
the hybrid exposure is influenced by several factors including  
the atmospheric conditions, the trigger status of all active detectors and their 
instantaneous data taking configuration. 
The calculation of the SD response is based on the deep knowledge of   
the array capability to trigger 
once a shower with a given energy and zenith angle hits the ground.  
Since the trigger in an EAS array is always a 
combination of trigger states of
neighboring detectors, the acceptance of any EAS array is directly connected to the probability 
that an individual detector triggers when a shower lands at a certain distance from it. 
This defines the concept of  ``Lateral Trigger Probability'' function.
This function has been used as a powerful tool for simulations in the analysis 
for the measurement of the hybrid energy spectrum~\cite{combined_spectrum} and 
of the atmospheric depth at shower maximum~\cite{xmax}.

The trigger design of the Auger surface detector is described in detail in~\cite{SD_acceptance}.
Each water Cherenkov detector of the surface array has a 10 m$^2$
water surface area and 1.2 m water depth, with three 9 in. photomultiplier tubes (PMTs) looking through optical coupling
material into the water volume, which is contained in a Tyvek\textsuperscript{\textregistered}
reflective liner. The signals provided by each PMT are digitised
by 40 MHz 10-bit Flash Analog to Digital Converters (FADCs)~\cite{engineering}. The achieved dynamical range
is sufficient to cover  with good precision both the
signals produced in the detectors near to the shower axis ($\sim$ 1000 particles/$\mu s$)
and those produced far from the shower
axis ($\sim$ 1 particle/$\mu s$). 
We recall here the basic structure of the used trigger algorithms. 
The two first levels (T1 and T2) 
are formed at each surface detector. Each trigger level can be divided in two modes, a threshold trigger (TH)
 and a time-over-threshold trigger (ToT). 
The first level threshold trigger (TH-T1) requires the coincidence 
 of the signals from the three PMTs equipping each station, each PMT signal being 
 above 1.75 ``Vertical Equivalent Muon'' (VEM)\footnote{ The distribution of measured light due to atmospheric muons produces a
peak in the PMT charge distribution, $Q^{\rm peak}_{\rm VEM}$
(or VEM in short), as well as
a peak in that of the pulse height, $I^{\rm peak}_{\rm VEM}$, both of them being
proportional to those produced by a vertical through-going muon~\cite{engineering}.}. 
The  TH-T1 trigger is used to reduce the rate due to atmospheric 
 muons to $\sim$100 Hz and can reach the second level, TH-T2, when the peak signal reaches at least 3.2 VEM 
 in coincidence between 3 PMTs signals, further reducing the rate to $\sim$ 20 Hz. The second mode, the ToT, requires 
 at least 13 time bins (i.e.\ more than 325 ns) in 120 FADC  bins of a sliding window of 3 $\mu$s to be above a threshold 
 of 0.2 VEM in coincidence in 2 out of 3 PMTs. 
 Time-over-threshold trigger stations are automatically promoted to the second level. 
 The threshold trigger is especially efficient at detecting strong narrow signals, mostly encountered 
 in horizontal showers or close to the axis of vertical showers. On the other hand, the ToT  
 is intended to select sequences of small signals spread in time.
 This is typical of low energy vertical showers dominated by an electromagnetic component
 or of high energy showers triggering stations at large distance from the shower axis because of muons produced 
 high in the atmosphere.

Higher level triggers are obtained by requiring the spatial and temporal coincidence of at least three stations satisfying the T2 conditions. In particular, for zenith angles below 60$^{\circ}$, the full efficiency for SD is reached at 10$^{18.5}$~eV~\cite{SD_acceptance}.  
In addition, if at least one FD telescope triggers in coincidence with one second level trigger station, a hybrid trigger is formed. 
Since FD has a lower energy threshold, hybrid events are 
also detected below the minimum energy for an independent SD trigger. 
For zenith angles below 60$^{\circ}$, the hybrid detector 
reaches nearly full efficiency at 10$^{18}$~eV~\cite{hybrid_aperture}.

In section 2, the  concept of a Lateral Trigger Probability (LTP) function 
is formalized and applied to the particular case of the 
surface detector of the Pierre Auger Observatory. 
In section 3, the LTP functions for a single time-over-threshold trigger station are derived and parametrized 
 for different primary particles (proton, iron, photon) and their dependence on energy and zenith 
 angle is explored for zenith angles up to 65$^\circ$
and for energies between $10^{17}$ and $10^{19}$~eV.
This energy range is relevant as it covers the interval in which the SD-only and the hybrid detection mode become fully efficient. 
The dependence on the choice of the hadronic interaction models is also discussed in section 3. 
In section 4, hybrid data are finally used to validate the simulation and to estimate    
the impact of weather conditions on the observed efficiency.
The LTP functions are found to reproduce very well the 
detector response over a wide range of energies and zenith angles.


\section{Concept of Lateral Trigger Probability}
\label{sec:concept}

The trigger probability of a single water Cherenkov detector depends on several independent physical parameters : i) the characteristics of the 
primary cosmic ray that initiates an air shower, e.g.,\ its energy and mass, ii) the type and geometry of the detector used to observe air 
showers (in the following we will only study water Cherenkov detectors used for the surface detector array of the 
Pierre Auger Observatory), iii) 
the trigger condition used to detect a signal from air showers   
iv) the geometry of the incoming shower, e.g.\, its incidence zenith angle and 
position with respect to the detector. 
To formalize these dependencies we define the Lateral Trigger Probability function  
$\Lambda_{A,E,\theta,Tr}(r,\phi)$ as the probability to trigger on an air shower induced by a primary particle of energy $E$, mass $A$ and 
zenith angle~$\theta$~\cite{ltpicrc2005}. 
Here, $r$ and $\phi$ are the radial coordinates of the single detector in the plane normal 
to the shower axis (shower frame).  Using a trigger condition $Tr$, this probability is simply given by:

\begin{equation}
         \Lambda_{A,E,\theta,Tr}(r,\phi) = \frac{N_{1}}{N_{1} + N_{0}}.
	 \label{eq:LTPDef2}
     \end{equation}
 where $N_{1}$ and $N_{0}$ are respectively the number of triggered and un-triggered detectors 
 with coordinates $r$ and $\phi$ in the shower frame.


\section{Simulations}
\label{sec:simulation}

The LTP functions have been derived using detailed
simulations of the EAS development and of the detector response.  
The simulation sample consists of about 15000 
CORSIKA~\cite{corsika} showers (proton, iron and photon primaries)  
with zenith angle distributed as 
$\sin\theta\cos\theta$ ($\theta < $65$^{\circ}$) 
and energies ranging between 10$^{17}$ and 
10$^{19}$~eV in steps of 0.25 in the logarithmic scale.
A ``thin sampling'' mechanism at the level of 10$^{-6}$ (optimal thinning) is applied following 
the standard method used for CORSIKA simulation with energies larger than 10$^{16}$~eV~\cite{thinning}.    
The showers have been generated    
with the models QGSJETII~\cite{qgsjet} and FLUKA~\cite{fluka} 
for high and low energy hadronic interactions.

In the simulation, the position of the shower core (i.e.\ the intersection of the shower axis with the ground) is  uniformly distributed over the surface array and  
each shower is used 5 times, each time  with a 
different core position, in order to cover different areas of the array and explore all the detector configurations. 
The surface detector response is simulated using GEANT4~\cite{geant4} and 
adopting the sampling procedure to regenerate particles in a ground detector
from thinned air shower simulations as described in~\cite{billoir}. 
The entire detector simulation is carried out  
within the framework provided 
by the Auger Offline software~\cite{offline}.

The trigger status of SD stations is inspected within a radius of 3 km from the shower axis  
and the Lateral Trigger Probability is then derived according to  
eq.~\ref{eq:LTPDef2}. At distances larger than 3~km, 
the trigger efficiency is negligibly small for the class of events studied in this paper.
All trigger modes of the surface detector are simulated in detail at all
 levels. However, for events with zenith angles below 65$^{\circ}$, the majority of the
stations forming a second level trigger
satisfy the ToT condition. In particular, for the considered zenith angles,  
the fraction of TH-T2 trigger stations not being also ToT is about 1\%,  
approximately independent of the energy. 
Thus, we focus the analysis on the ToT stations.\\  
\begin{figure}[p]  
  \begin{center}
    \includegraphics[width=0.49\textwidth]{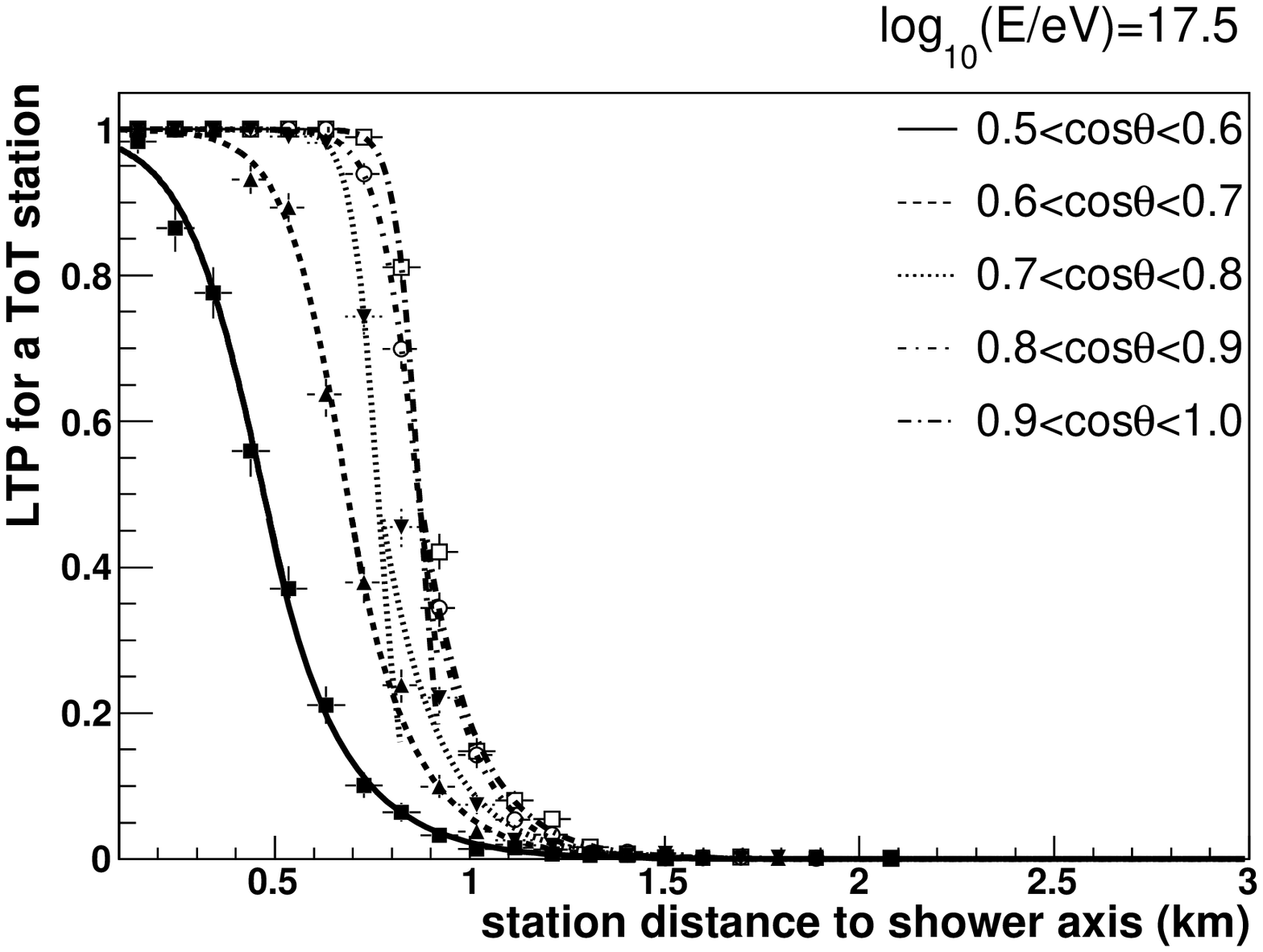} 
    \includegraphics[width=0.49\textwidth]{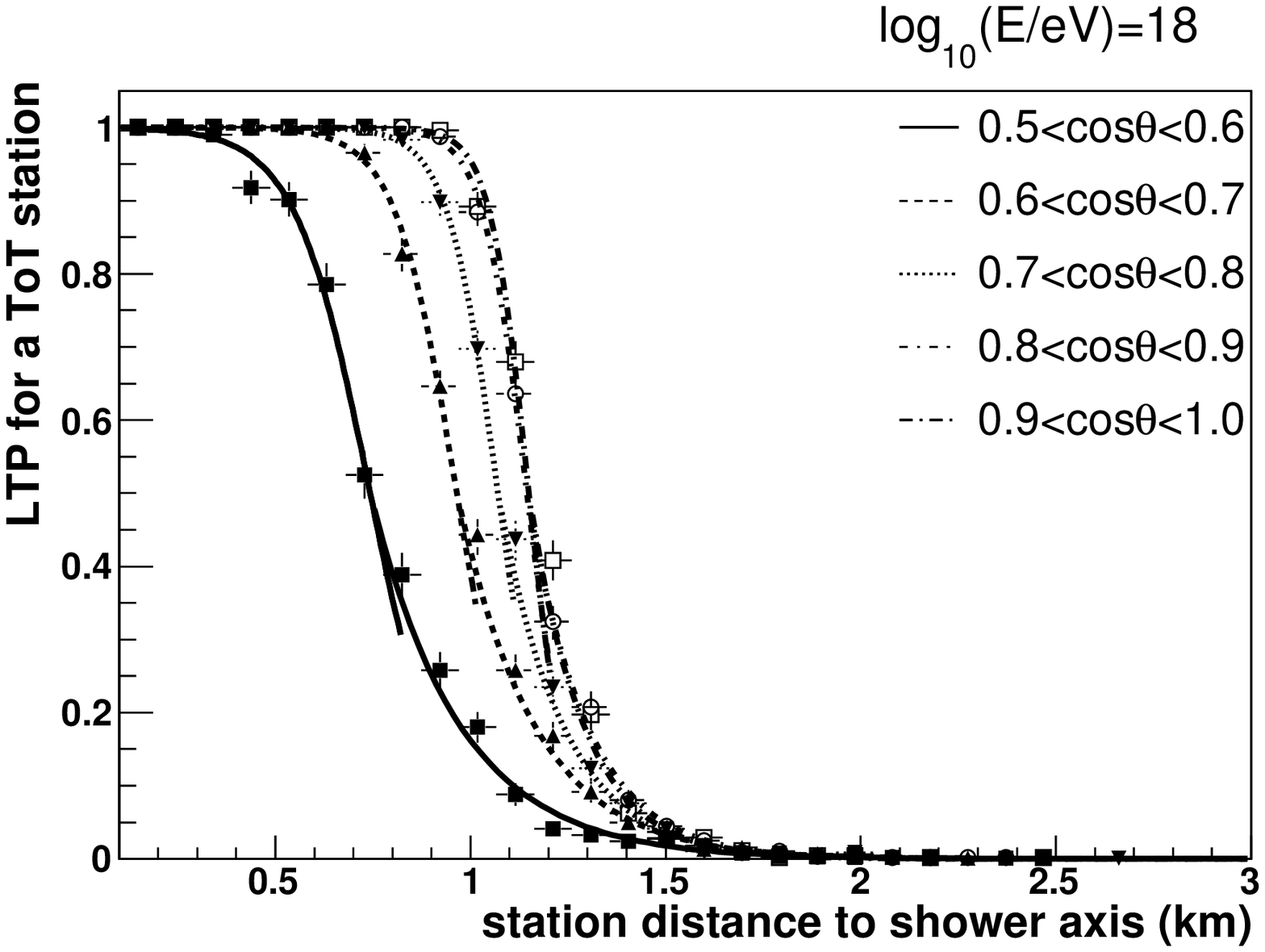} 
    \vskip 0.5 cm
    \includegraphics[width=0.49\textwidth]{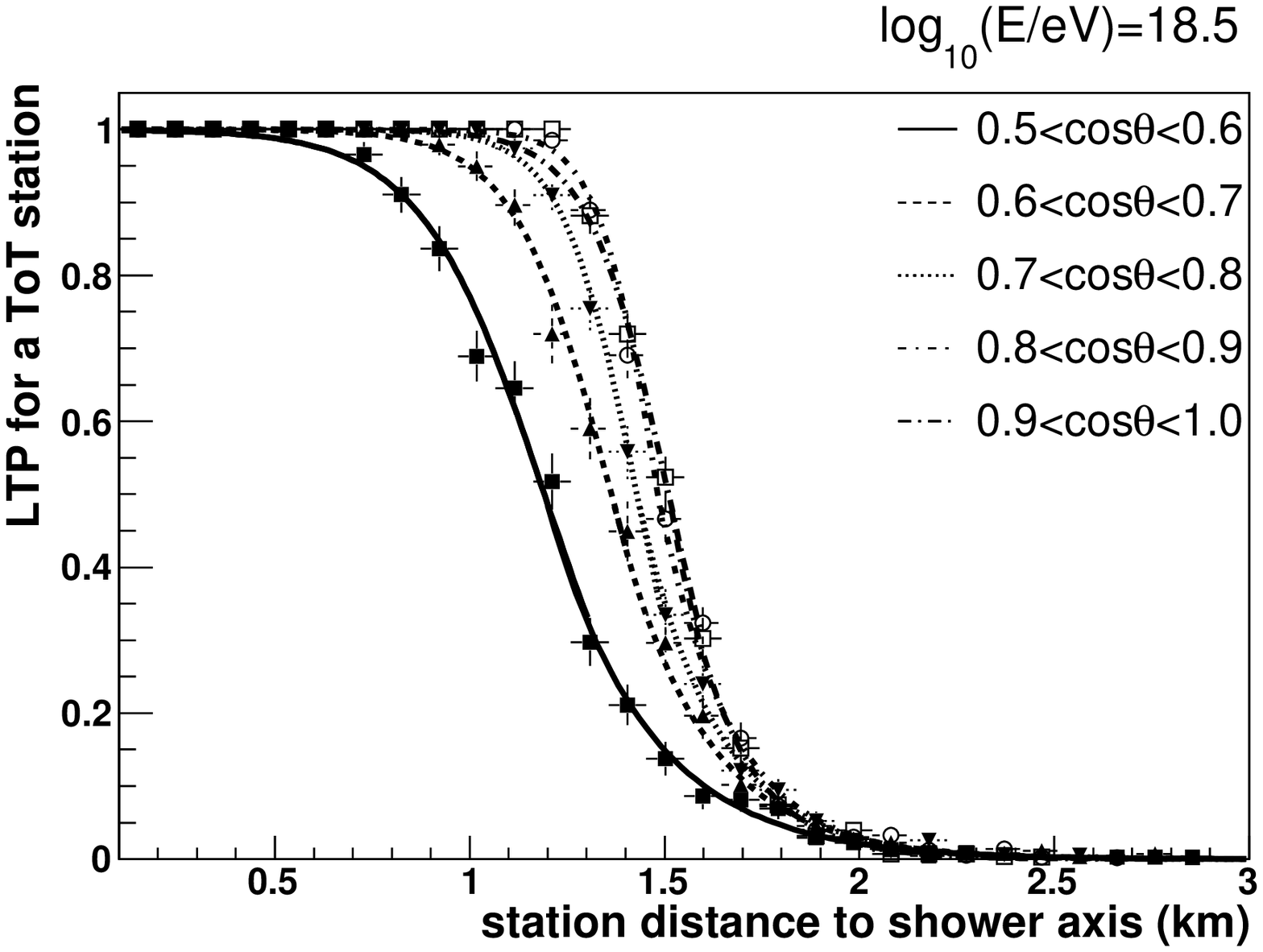} 
    \includegraphics[width=0.49\textwidth]{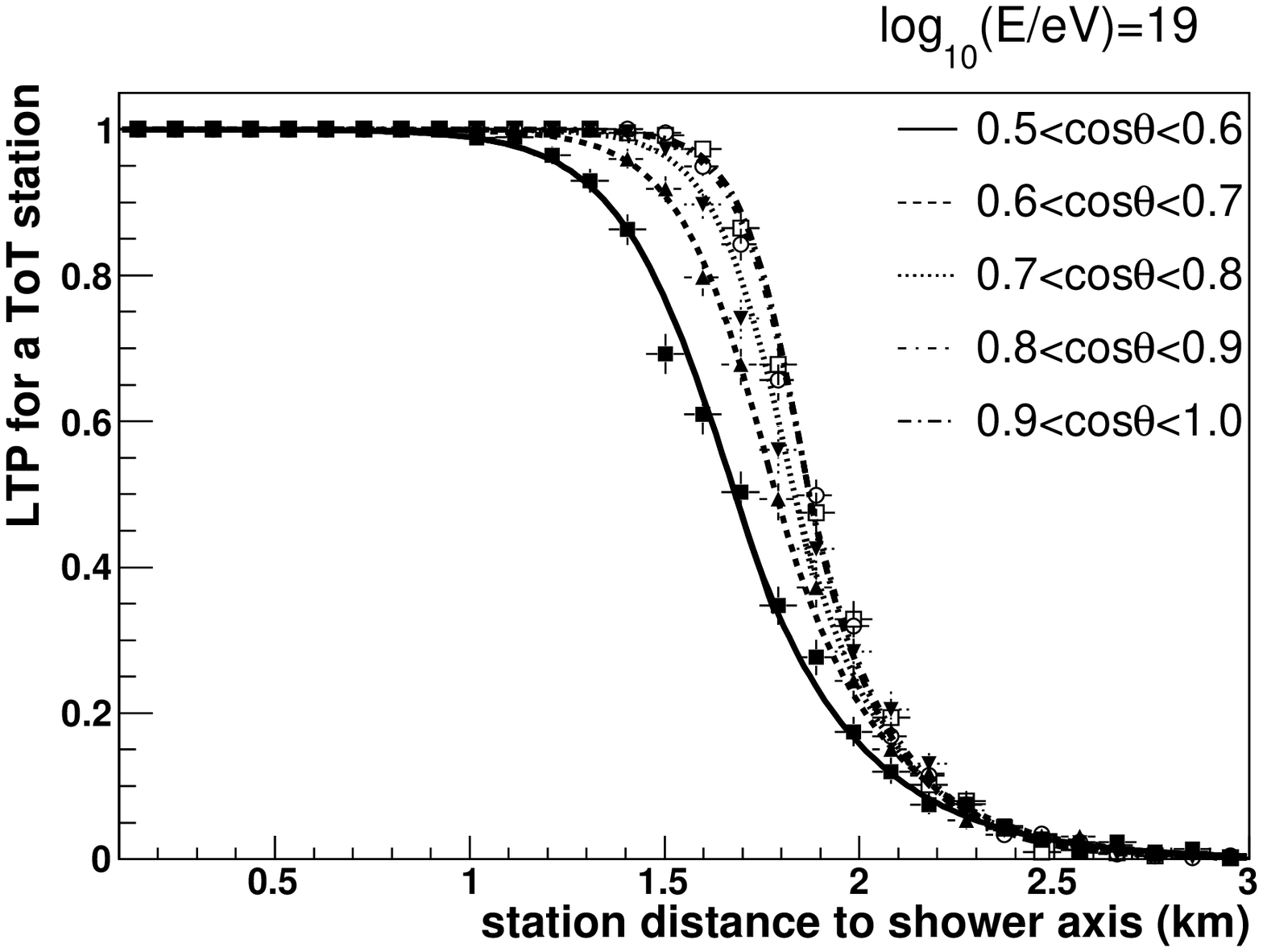} 
    \caption{Lateral Trigger Probability from simulations (proton primary) for a ToT station at a 
given energy, from 10$^{17}$~eV up to 
      10$^{19}$~eV in steps of 0.5 in the logarithmic scale.  
Different bins of $\cos\,\theta$ are also shown together with a fit performed according to 
eq.~\ref{eq:LTP},  superimposed as a continuous line. }    
    \label{fig:LTP_zen}
    \vspace{0.6cm}
    \includegraphics[width=0.49\textwidth]{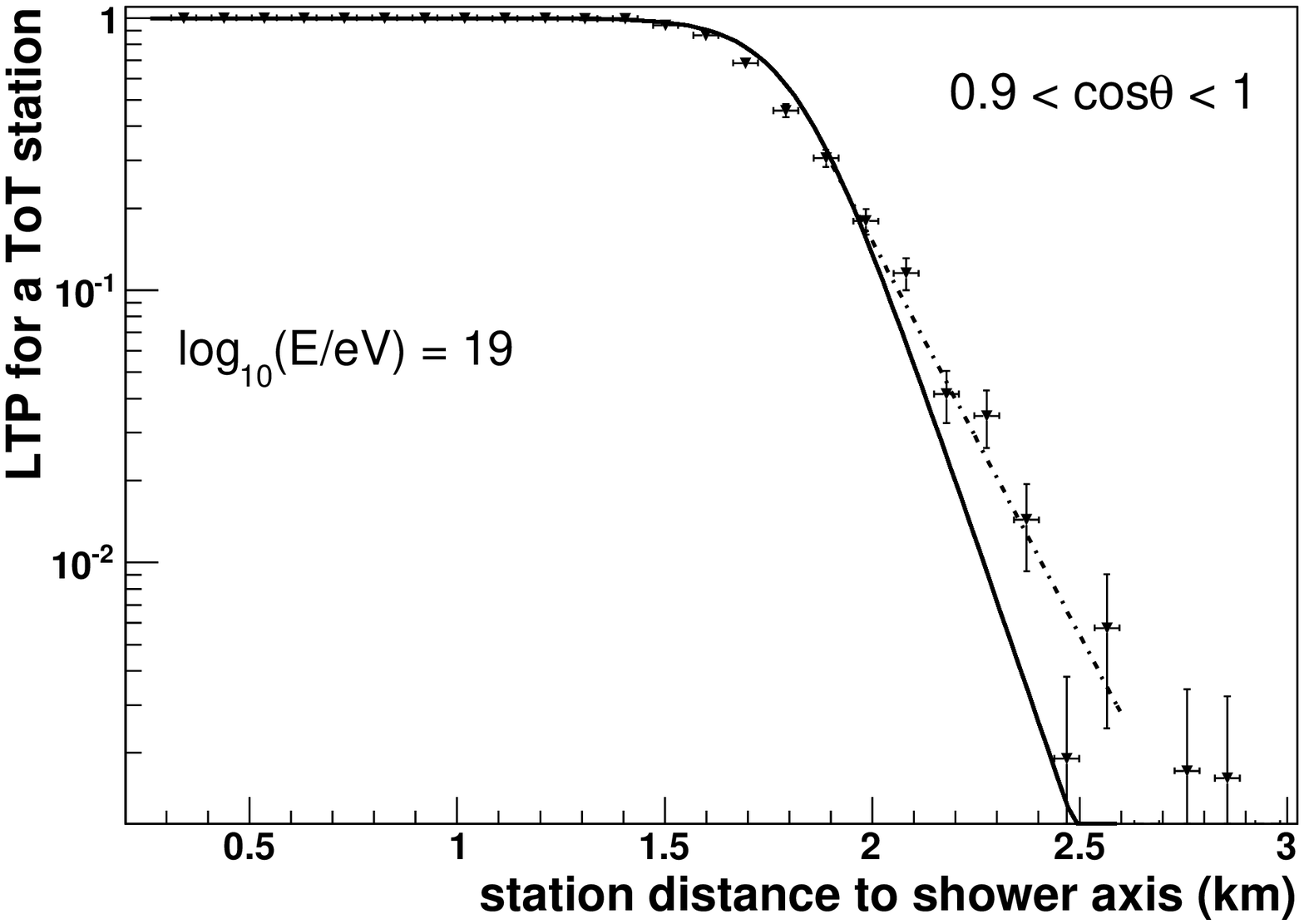}
    \includegraphics[width=0.49\textwidth]{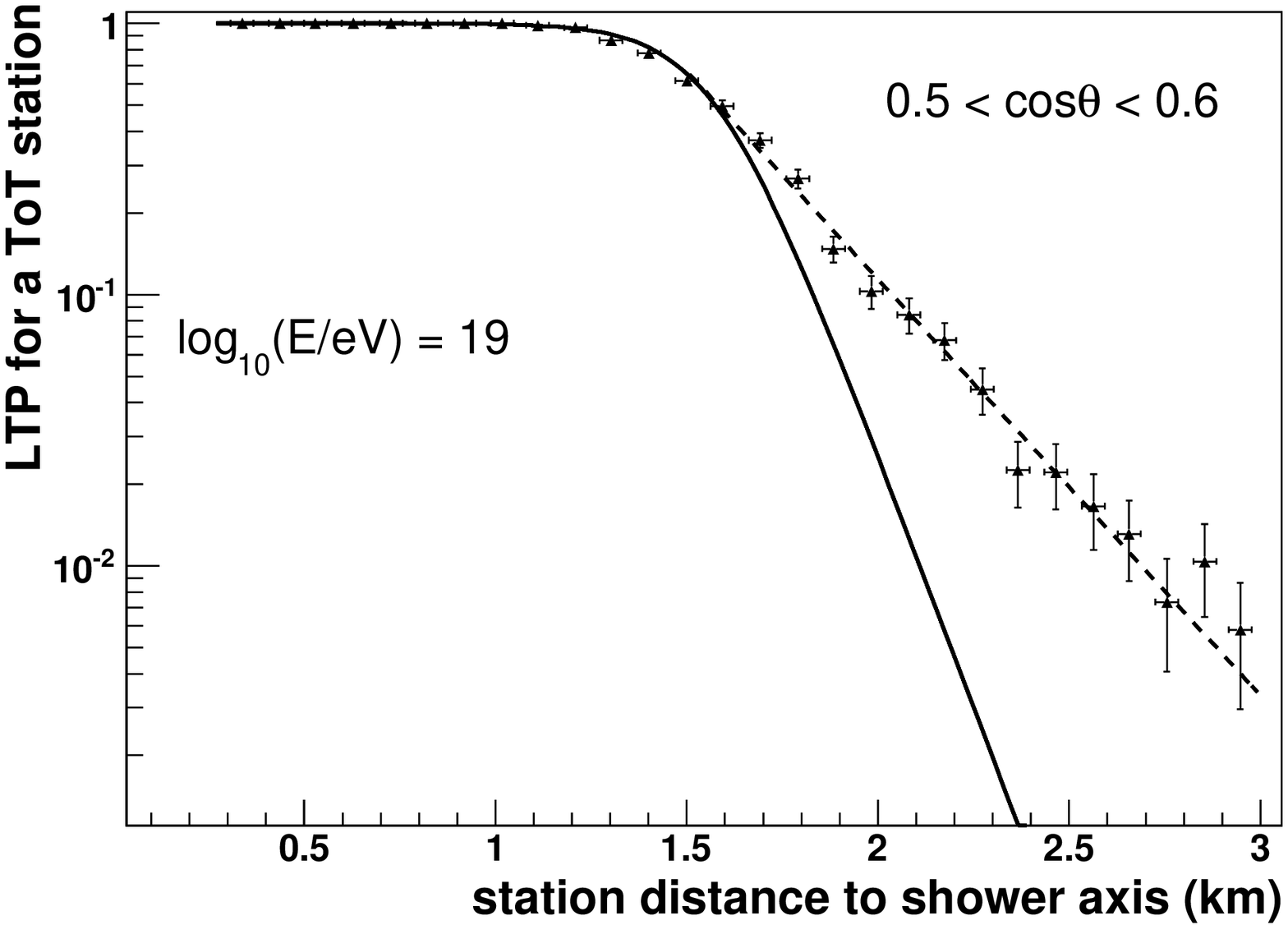} 
    \caption{Fit made with a step function in proximity of the shower axis (continuous line) 
 and by an exponential at larger distances (dashed line). The ToT probability 
is shown for    
 vertical (left) and inclined (right) showers at energy of 
10$^{19}$~eV.}    
    \label{fig:sigmoexpo}
  \end{center}
\end{figure}
The lateral trigger probability for a ToT station 
 is  shown in Fig.~\ref{fig:LTP_zen} at a given energy and for different ranges of 
the cosine of the zenith angle~$\theta$.    
The maximum effective distance for detection increases with 
energy and, for a given energy, with the cosine of the zenith angle, i.e.\   
events with larger zenith angle tend to trigger less due to the attenuation of their electromagnetic component.
For moderately inclined showers, an asymmetry is expected in the signal detected in
the stations placed at the same distance to the shower axis but with different azimuth in the shower
frame~\cite{dova_asym}. Indeed, secondary particles arriving earlier
traverse less atmosphere and are less attenuated than the late ones.  
As a consequence, 
early stations may exhibit larger trigger probabilities
and produce larger signals. 
Actually, for zenith angles below 65$^{\circ}$, this effect has been found to have 
 a quite low influence on the trigger probability, only 
 noticeable
above 30$^{\circ}$ (in simulations as
well as in the data).
In the following we consider LTP functions averaged over all azimuths in the showers frame. 
A more detailed treatment including the azimuthal
dependence does not introduce measurable differences for acceptance
calculations.

\begin{figure}[p]  
  \begin{center}
    \includegraphics[width=0.9\textwidth]{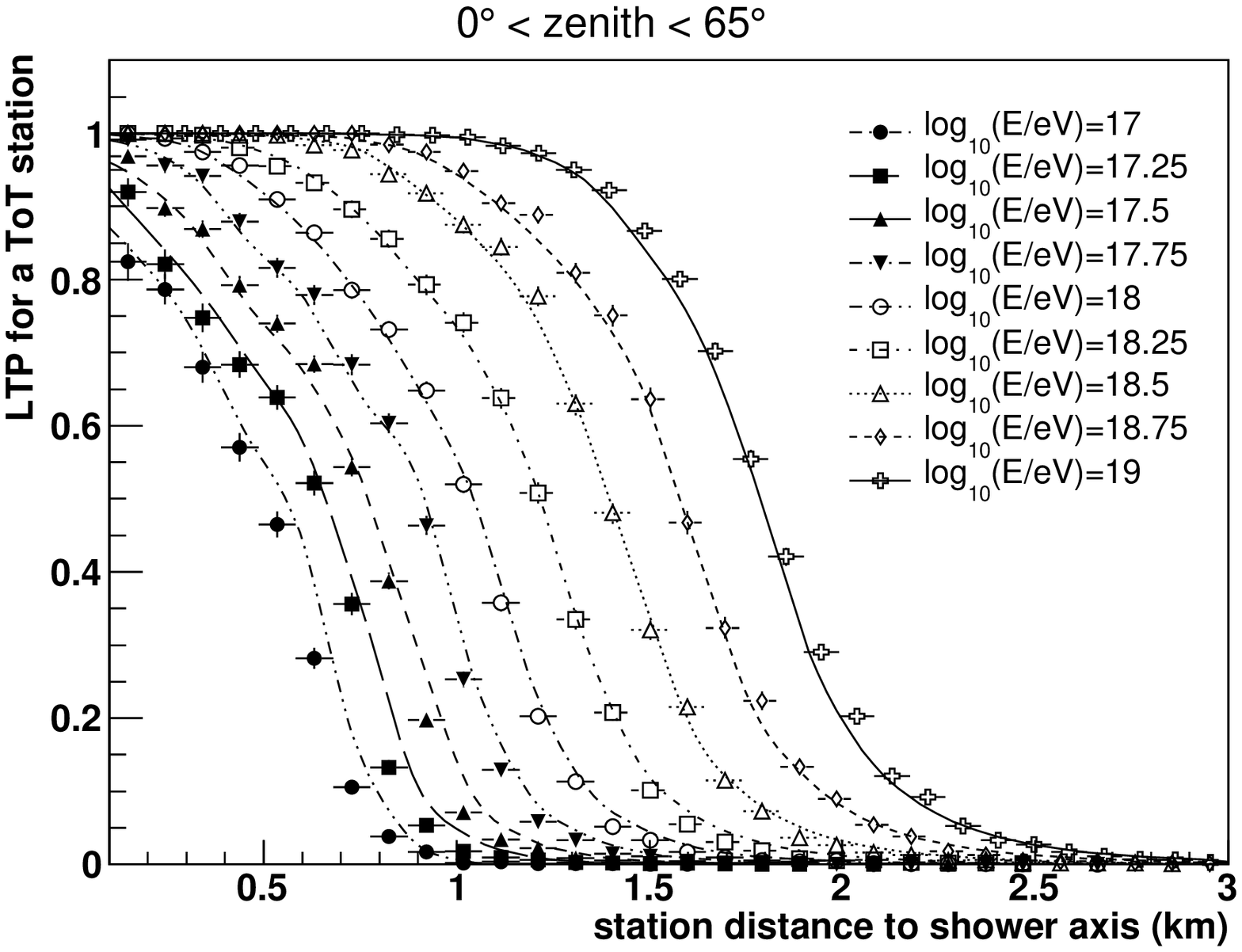}
    \caption{Lateral Trigger Probability for a ToT station   
as a function of station distance to shower axis and for different energies 
 (proton primary). The outcome of the parametrization 
  is superimposed as a line. All zenith angles up 
to 65$^{\circ}$ are merged.}    
    \label{fig:param}
  \end{center}
\end{figure}

A fit combining a step function (close to the axis) with   
an exponential (further away) reproduces reasonably well 
the full simulated data set.  
The form of the fit function used is:
\begin{equation}\label{eq:LTP}
\text{LTP(r)} = \left\{ \begin{array}{ll}
\frac{1}{1+e^{-\frac{r-R_0}{\Delta R}}} & \textrm{$r\leq R_0$}\\ \\
\frac{1}{2}e^{C\cdot (r-R_0)} & \textrm{$r> R_0$}\\
\end{array} \right.
\end{equation}
where $R_{0}$, $\Delta R$ and  $C$ are free fit parameters, 
with $R_{0}$ being the distance where LTP is
equal to 0.5. 
A fit performed according to eq.~\ref{eq:LTP} is superimposed on each 
plot shown in Fig.~\ref{fig:LTP_zen}.
As an example, the ToT trigger probability at energy $E$=10$^{19}$~eV and 
for two angular bins (vertical showers on the left 
 and showers with larger zenith angle on the right) is shown in Fig.~\ref{fig:sigmoexpo}: 
 the exponential can reproduce very well the tail of the probability 
distribution at large distances from axis,
 in particular for inclined events. 
The dependences of fit parameters  $R_{0}$, $\Delta R$ and  $C$  
on energy and zenith angle can be parametrized by quadratic polynomials    
 in the variables $\cos\,\theta$ and $\log_{10}(E/eV)$. 
The corresponding coefficients are tabulated in the Appendix for proton, iron and photon primaries.
In Fig.~\ref{fig:param}, 
the ToT trigger probability from parametrization  
has been superimposed on the simulation (proton primary, all zenith angles up to 65$^{\circ}$ are merged).
The comparison is performed as in the following. 
For each simulated event, i.e.\ for a certain primary, energy and arrival direction, the LTP is calculated using the parametrization (lines) and 
shown together with the full simulation (points).  
The agreement is remarkably good in the entire energy range for proton (shown in the figure) and for   
iron and photon primaries.     

\begin{figure}[p]  
  \begin{center}
  \vspace{-1.5cm}
    \includegraphics[width=0.57\textwidth]{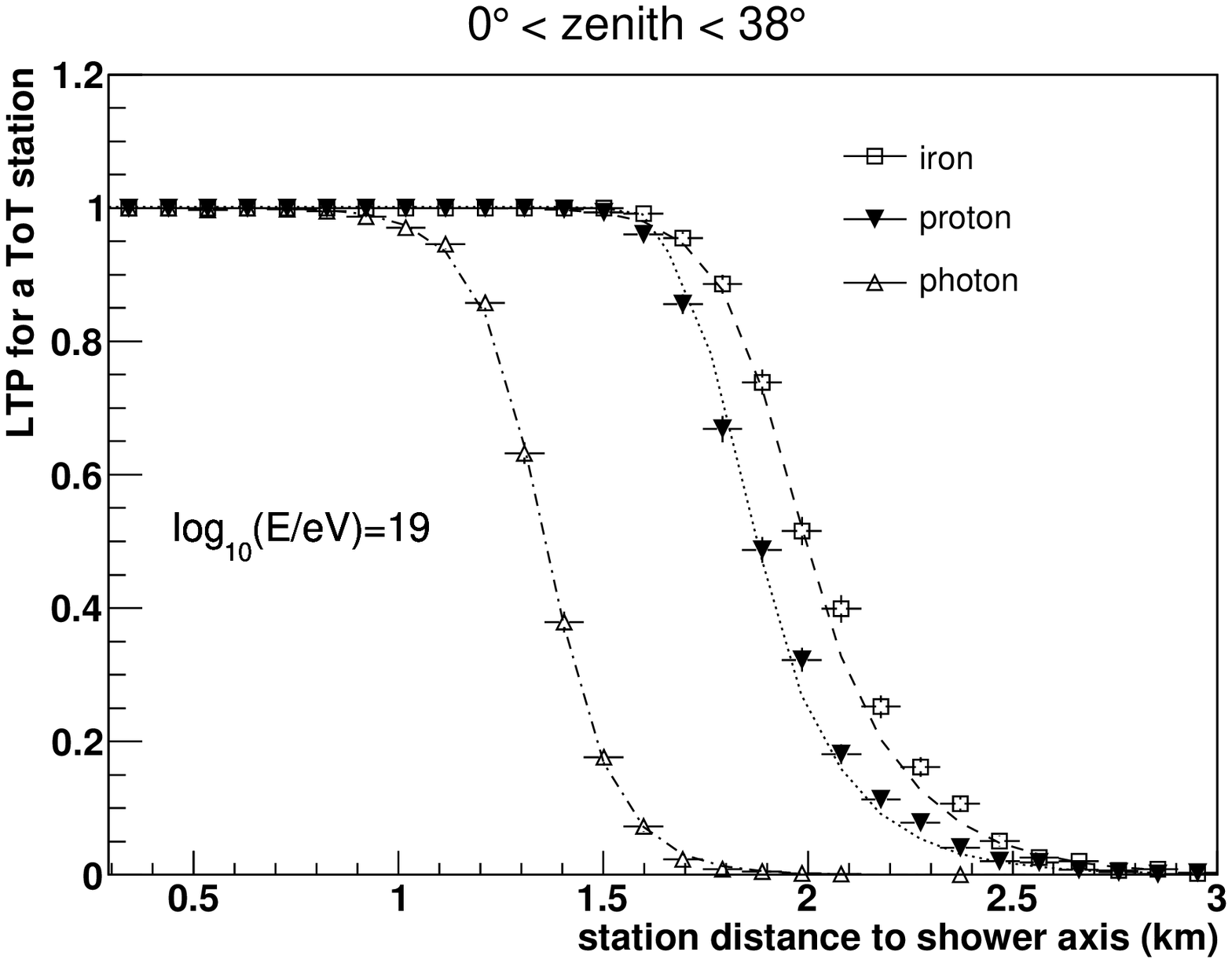}
    \vskip 0.3 cm
    \includegraphics[width=0.57\textwidth]{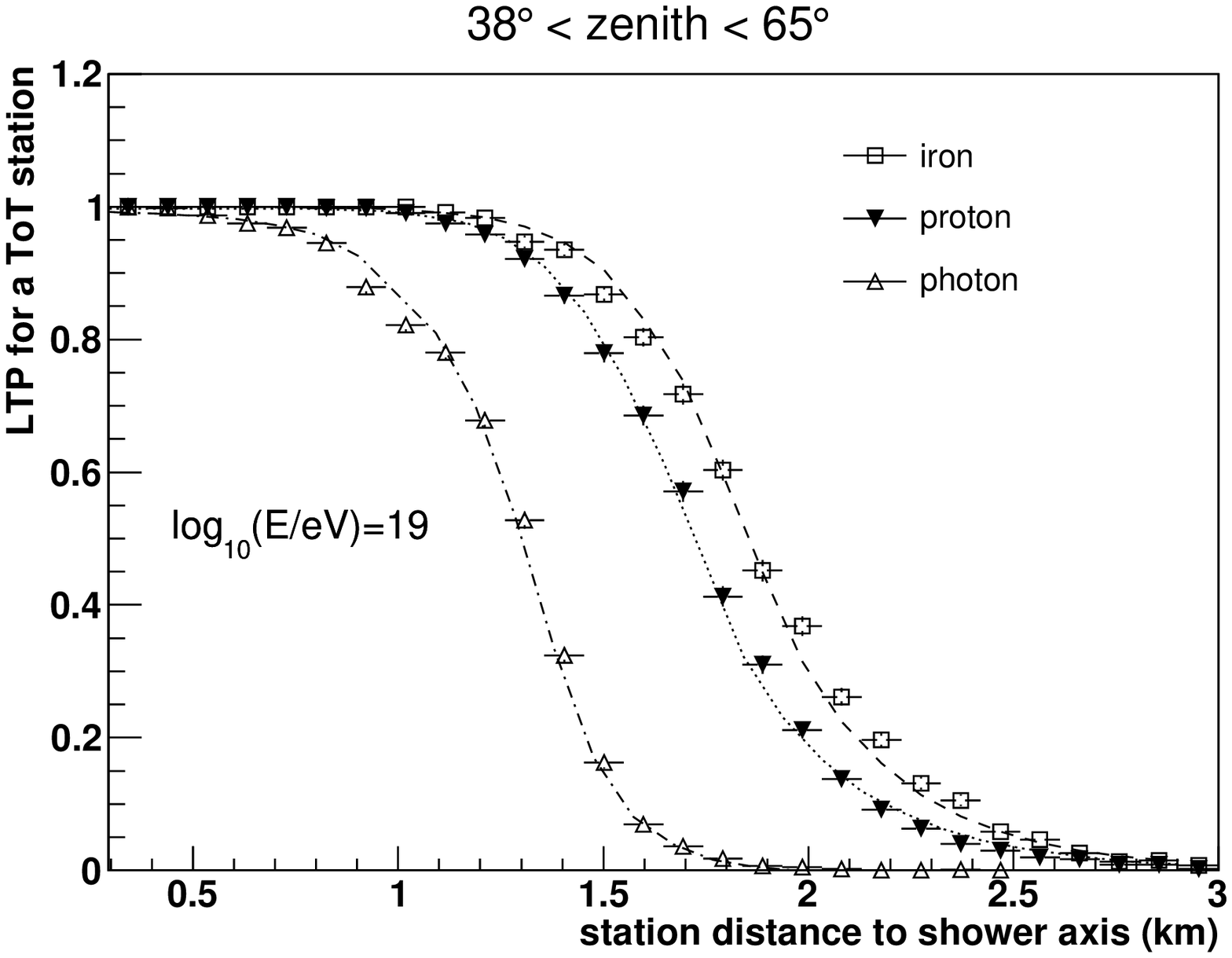}    
    \caption{Lateral Trigger Probability for a ToT station. Proton, iron and photon primaries of energy 10$^{19}$~eV for 
    two zenith angle ranges,
   0$^{\circ}$~-~38$^{\circ}$ (top) and 38$^{\circ}$~-~65$^{\circ}$ (bottom). The outcome of the 
    parametrization is superimposed as a continuous line.}    
    \label{fig:mass}
    \includegraphics[width=0.8\textwidth]{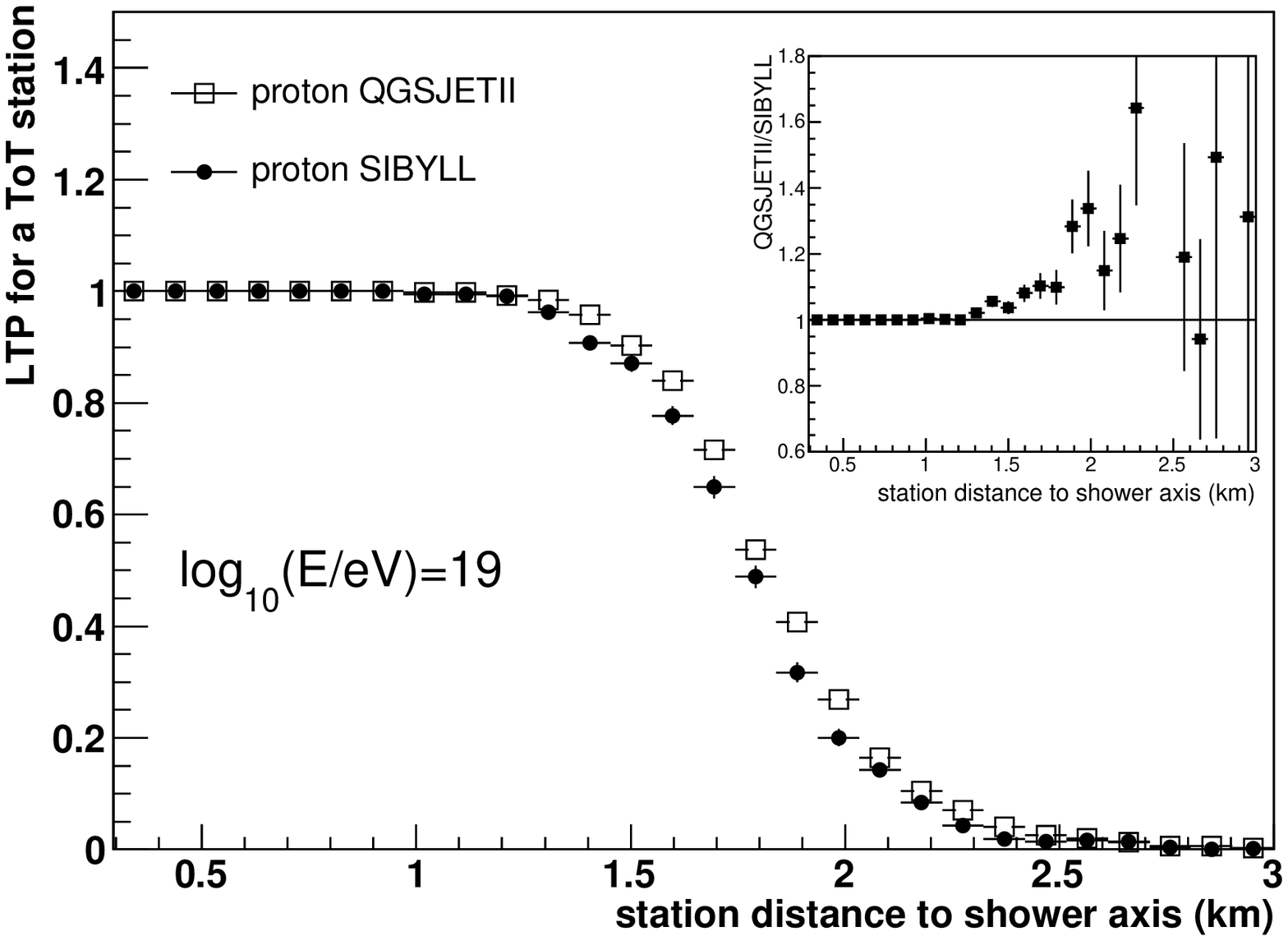}
    \caption{Lateral Trigger Probability for a ToT station (zenith angle between 0$^{\circ}$ and 65$^{\circ}$). 
    Proton primary at energy of 10$^{19}$~eV with QGSJETII and SIBYLL.
    The ratio QGSJETII/SIBYLL is shown in the inset. }    
    \label{fig:model}
  \end{center}
\end{figure}

\subsection{Dependence on primary mass}
The detector response to showers induced by different primary particles is shown 
in Fig.~\ref{fig:mass}, for two classes of events, vertical ($0^{\circ}<\theta< 38^{\circ}$) on the top 
and moderately inclined ($38^{\circ}<\theta<65^{\circ}$) on the bottom.
Because of their larger number of muons, showers induced by iron nuclei provide a 
higher trigger capability at larger distances than those induced by protons,  
for all zenith angles.
However, the difference between proton and iron is too small to give any hint for mass composition analysis.
On the other hand, the LTP functions for photon primaries differ sensibly from those of hadrons   
(they vanish at shorter distances, about 500 m less at an energy of 10$^{19}$~eV). 
This is a consequence of the structure of the lateral distribution of photon showers, i.e.\ at a
given energy, their effective {\em footprint} at the ground is smaller than the one of hadrons. 
Moreover, in photon showers  
there is a much smaller number of muons. 

It is worth noting that the energy threshold corresponding to full efficiency for SD, derived 
from data and simulation
in ref.~\cite{SD_acceptance}, has been found to be compatible 
with the expectation for hadronic primaries.

\subsection{Dependence on hadronic interaction model}
Different choices of high energy interaction models
influence the simulation of shower development and could affect the expected trigger efficiency.  
The dependence of the Lateral Trigger Probability on the assumptions for the hadronic interaction model 
has been investigated using a sample of simulated showers (proton) produced with SIBYLL~\cite{sibyll}.  
As shown in Fig.~\ref{fig:model}, the LTP functions derived with the 
two hadronic interaction models differ only at large distance from the shower axis, in a range where 
the efficiency degrades rapidly.  In this region, 
SIBYLL gives a lower LTP since  
this model predicts on average a smaller number of muons.  
 Those differences are however too small to imply an observable impact on the detector acceptance.

\section{LTP functions from data and comparison with simulation}
\label{sec:data}

The LTP functions can be derived from data by calculating the 
ratio of triggered to active stations 
within a given distance from the reconstructed shower axis.
While doing this, the actual surface detector configuration must   
be accurately taken into account as a function of time.   
In addition, only high quality data are selected to avoid biases due to mis-reconstructed energies and/or arrival directions.
The use of hybrid events allows to derive LTP functions 
also for energies below the threshold of an independent SD trigger.
This is a benefit of the hybrid design that aims 
to fully exploit the distinctive potential offered by the Pierre Auger Observatory.
 Two years of hybrid data collected between June 2006 and May 2008 were used for this study. The events are selected  
 as described in~\cite{hybrid_aperture} and this ensures an angular resolution of about 0.6$^\circ$ and a core 
 position determination better than 70~m. Further requirements on the goodness of the reconstructed 
longitudinal profile provide an energy  resolution of about 10\% above 10$^{18}$~eV and less than 15\% at 
lower energies~\cite{hybrid_aperture}. 

\begin{figure}[t]
    \includegraphics[width=0.52\textwidth]{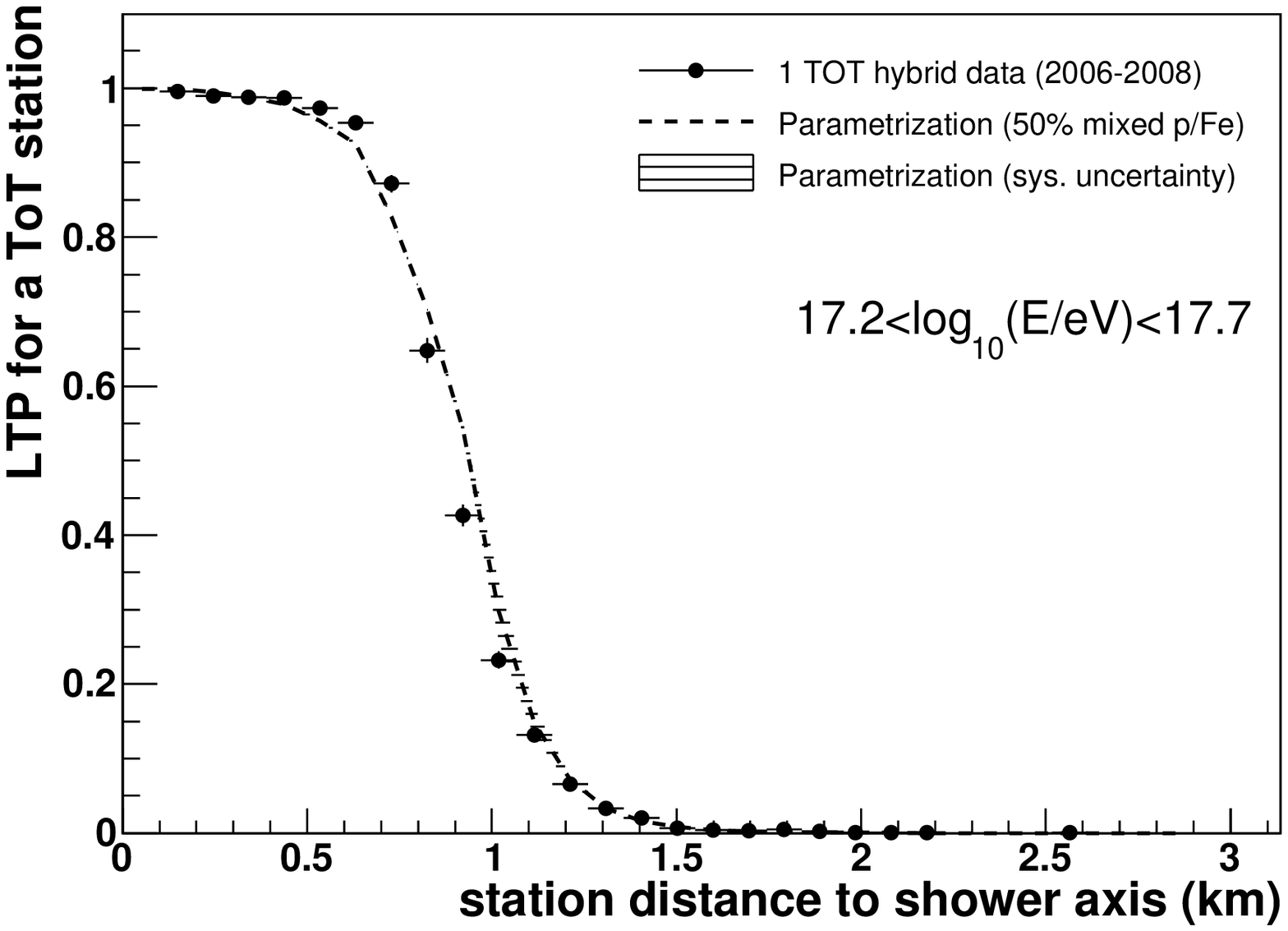}
    \hspace{-.5cm}
    \includegraphics[width=0.52\textwidth]{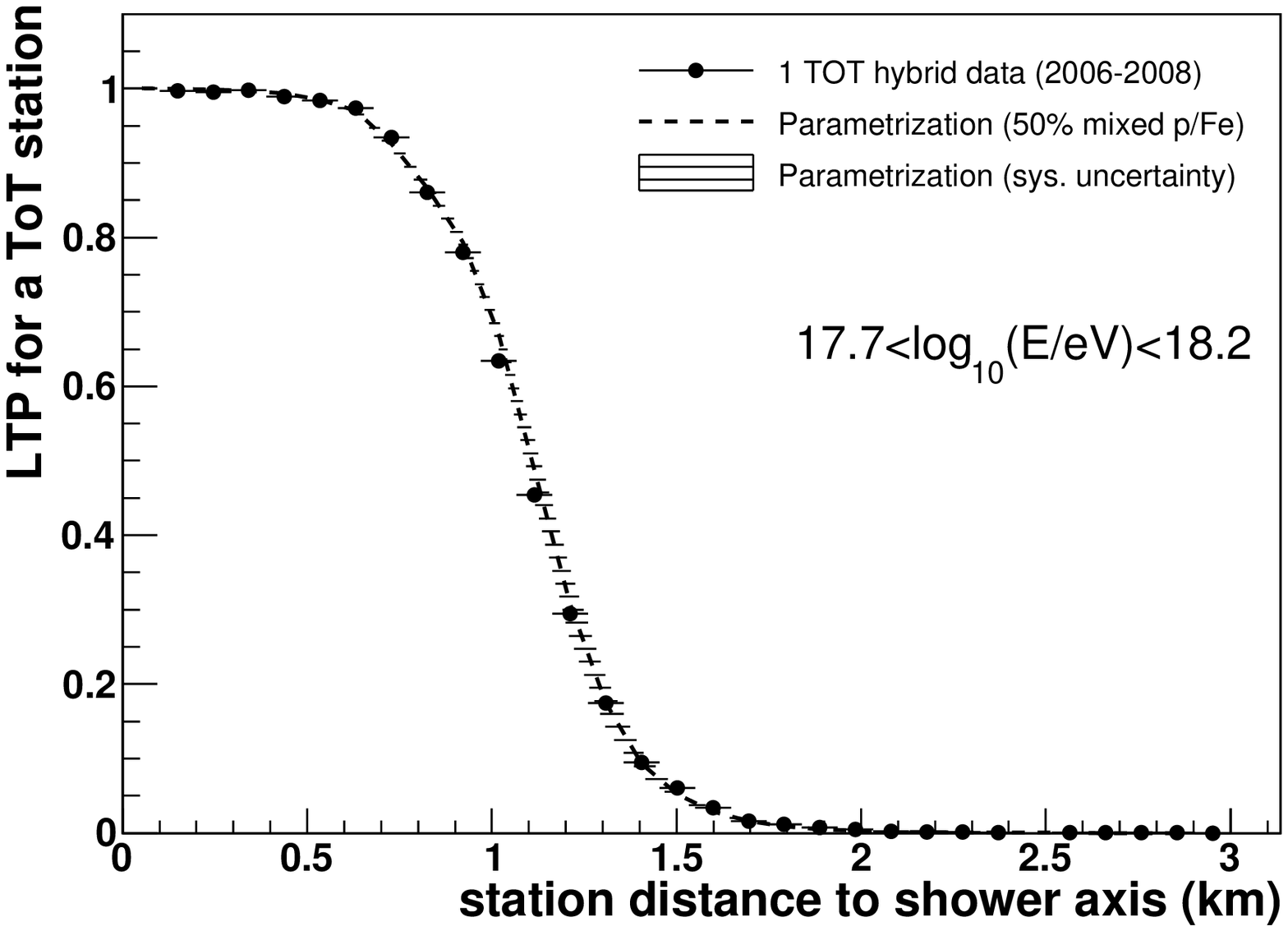}
    \vskip 0.3 cm 
    \includegraphics[width=0.52\textwidth]{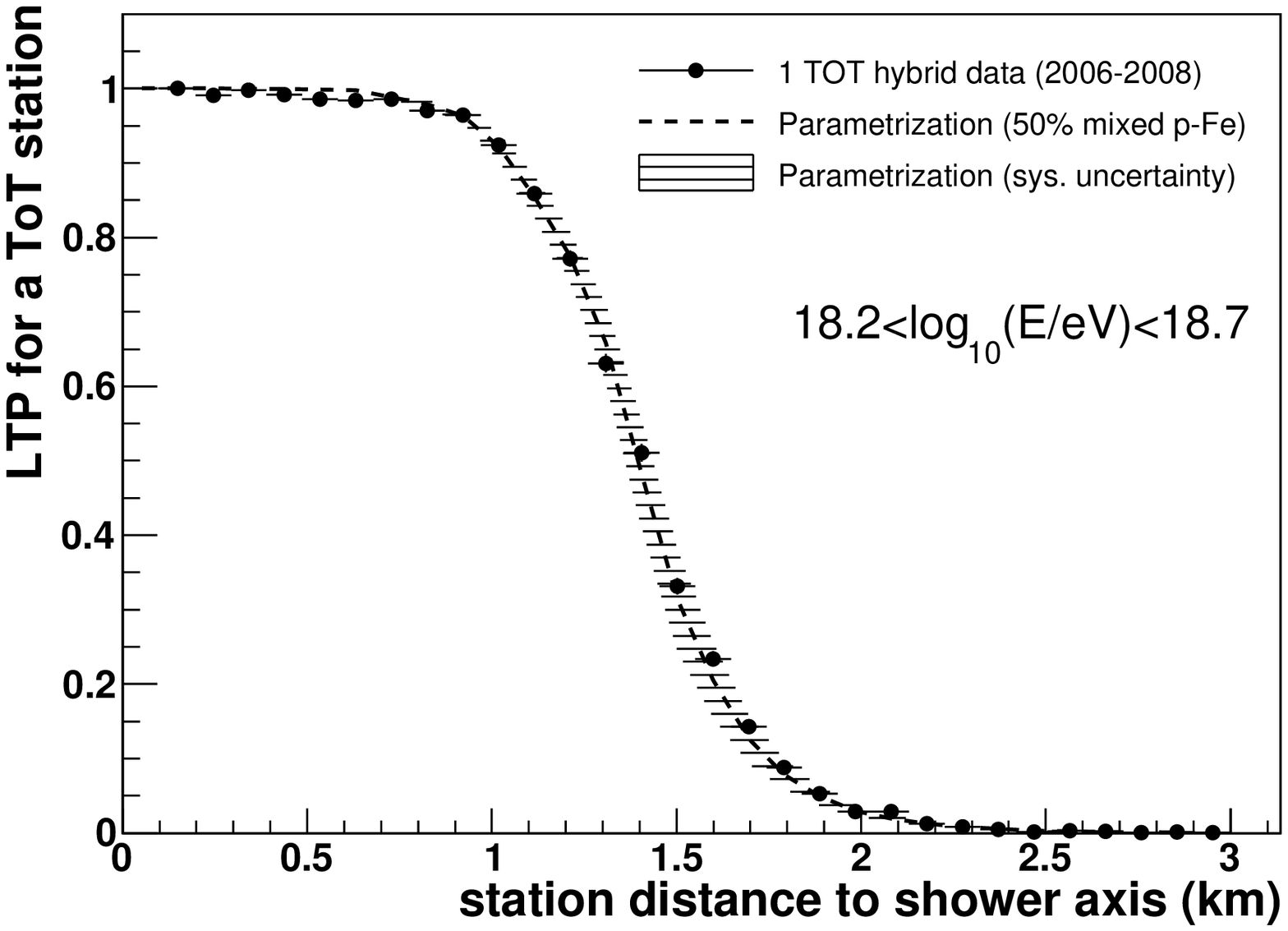}
    \hspace{-.5cm}
    \includegraphics[width=0.52\textwidth]{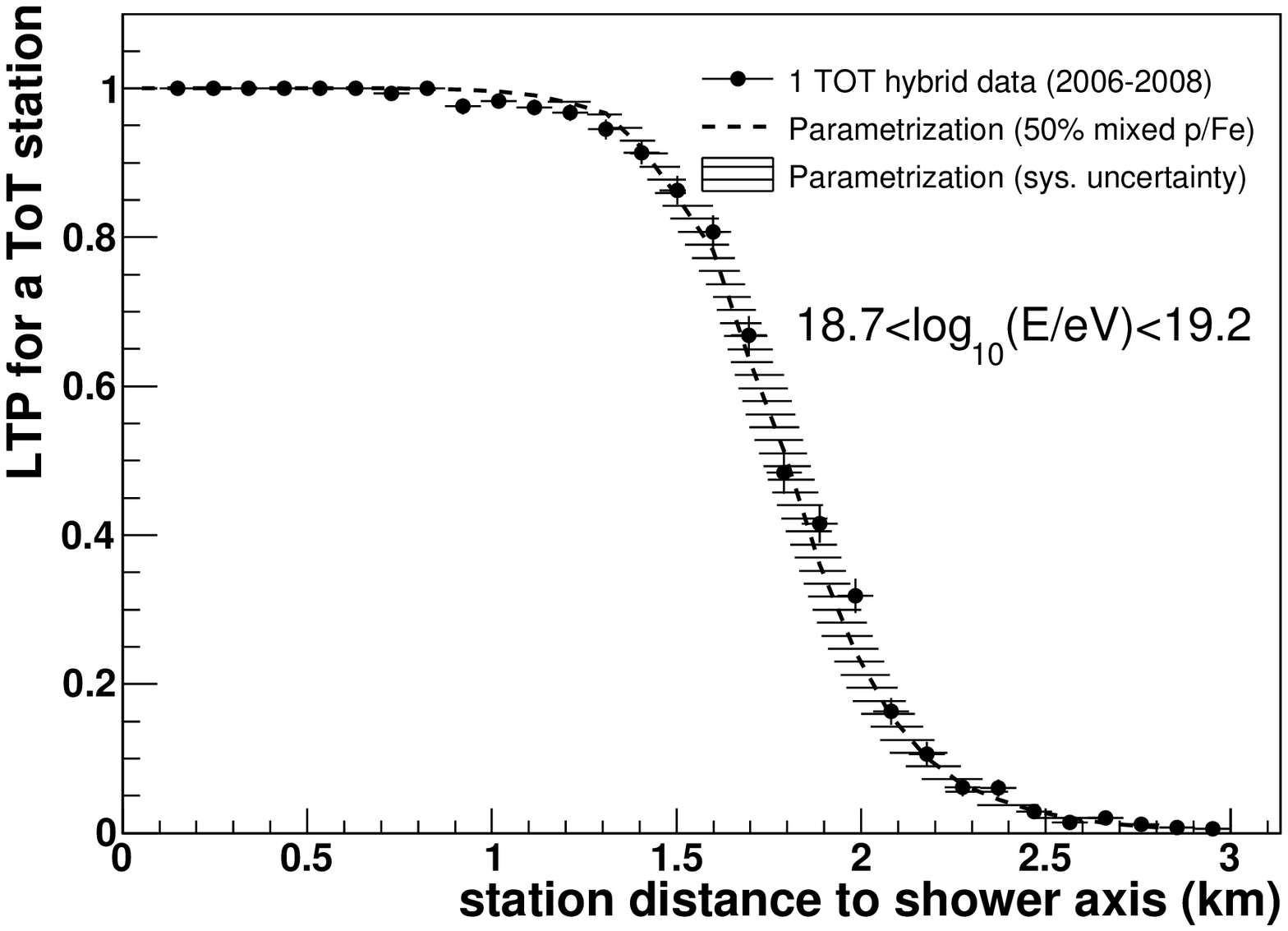}        
\caption{Comparison of simulation with hybrid data collected 
in two years. All zenith angles up to 65$^{\circ}$ merged.
The energy intervals are 
10$^{17.2}$ $<$ $E$ $<$ 10$^{17.7}$~eV,   
10$^{17.7}$ $<$ $E$ $<$ 10$^{18.2}$~eV, 
10$^{18.2}$ $<$ $E$ $<$ 10$^{18.7}$~eV, 
10$^{18.7}$ $<$ $E$ $<$ 10$^{19.2}$~eV.
}
\label{fig:sim-data}
\end{figure}

The LTP measured from data is shown in Fig.~\ref{fig:sim-data} for different energy intervals.
To verify the performance of the parametrization described in section~\ref{sec:simulation}, for each selected event,  
the LTP of any    
active station within 3~km from the shower axis is calculated using the reconstructed 
energy and direction. The predicted probability (dashed line) is then superimposed 
on data (points), see Fig.~\ref{fig:sim-data}. 
In this way, data are compared to simulation taking into account the actual status of the detector.
The shaded area gives the interval of expected values assuming that data are pure proton (lower edge) or pure iron (upper edge).
A 50\% proton and 50\% iron mixed composition has been assumed for the parametrization (dashed line).
The agreement is  good over the entire energy range. This feature actually starts   
 at very low energies, even below the range of full efficiency for the hybrid detection~\cite{hybrid_aperture}. 
 In this case, whereas in data only events with at least one SD ToT station are selected, 
 in simulation also the events that did not trigger at all are taken into account in the calculation of the probability. 
 As a consequence, the comparison between data and simulation  could be biased. However, the good level of agreement actually reached 
 reflects the fact that the hybrid detection is very close to fully efficient 
 and the energy reconstruction remains reasonably good within the scope of this analysis down to energy of about 10$^{17.5}$~eV.

\begin{figure}[p]
\vspace{-2.7cm}
\begin{center}
    \includegraphics[width=0.97\textwidth]{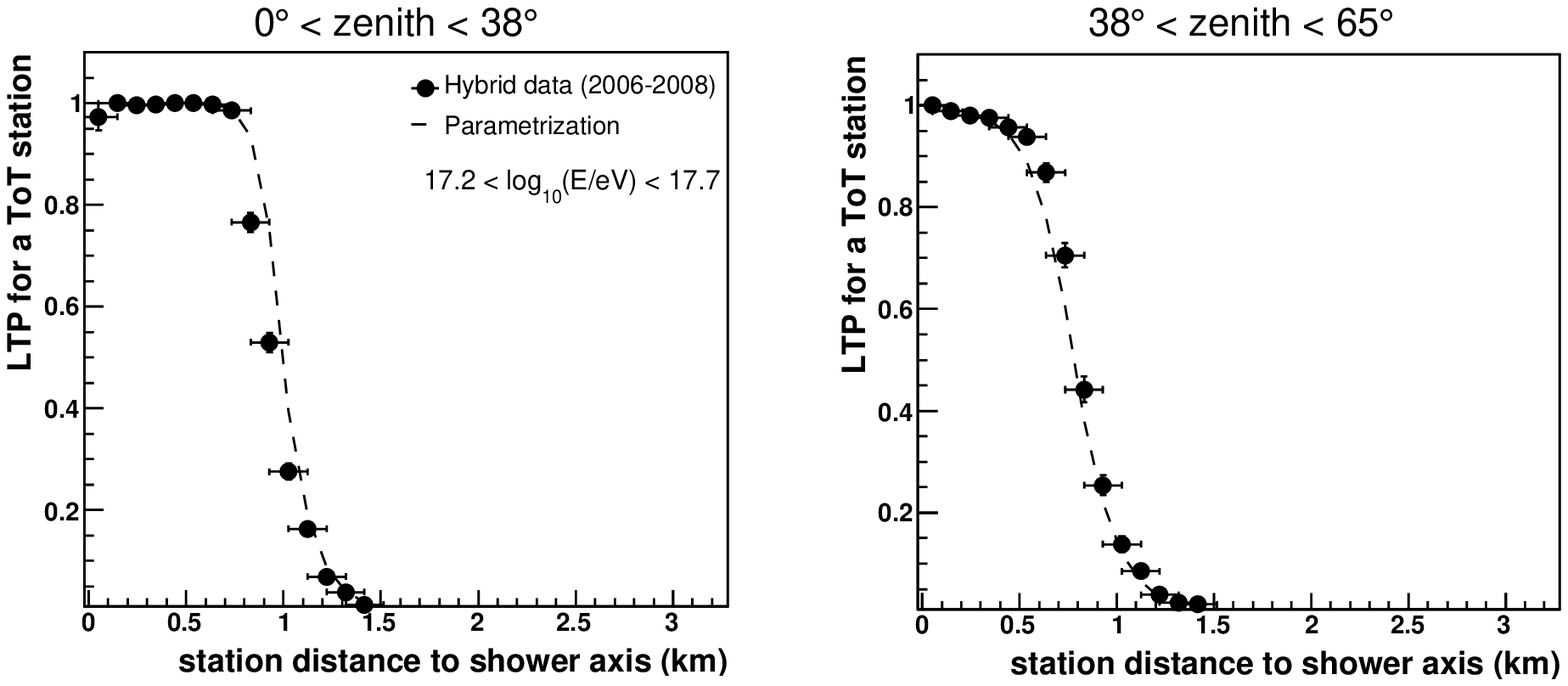}
    \vskip 0.17 cm 
    \includegraphics[width=0.97\textwidth]{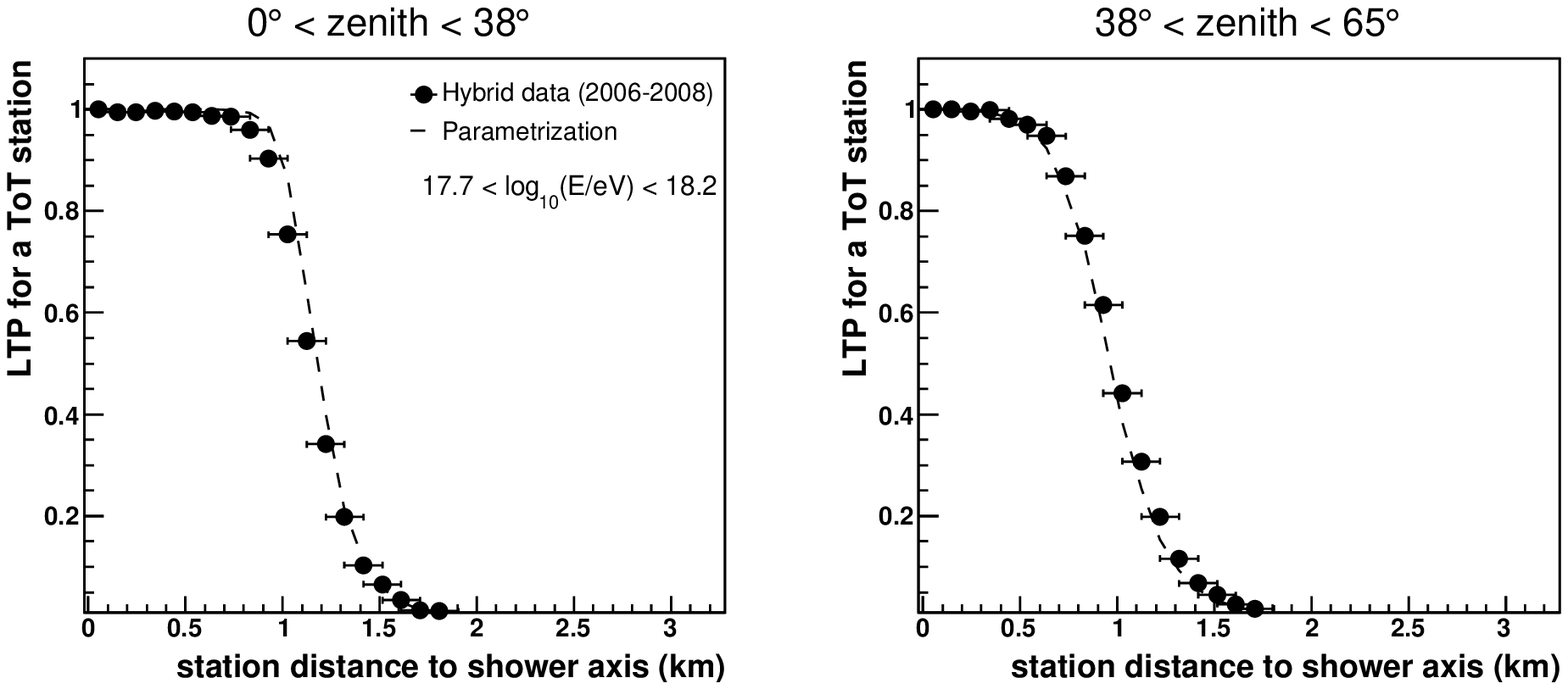}
    \vskip 0.17 cm     
    \includegraphics[width=0.97\textwidth]{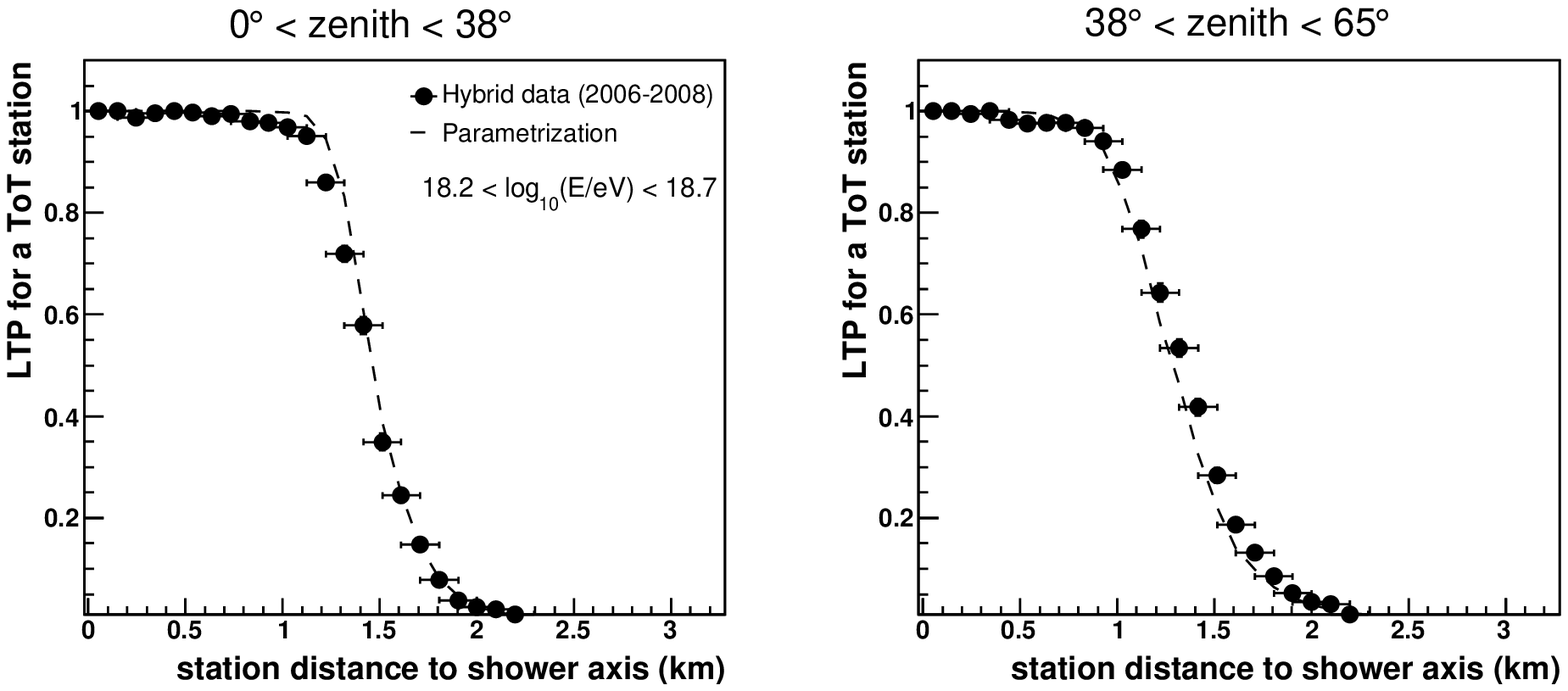}
    \vskip 0.17 cm     
    \includegraphics[width=0.97\textwidth]{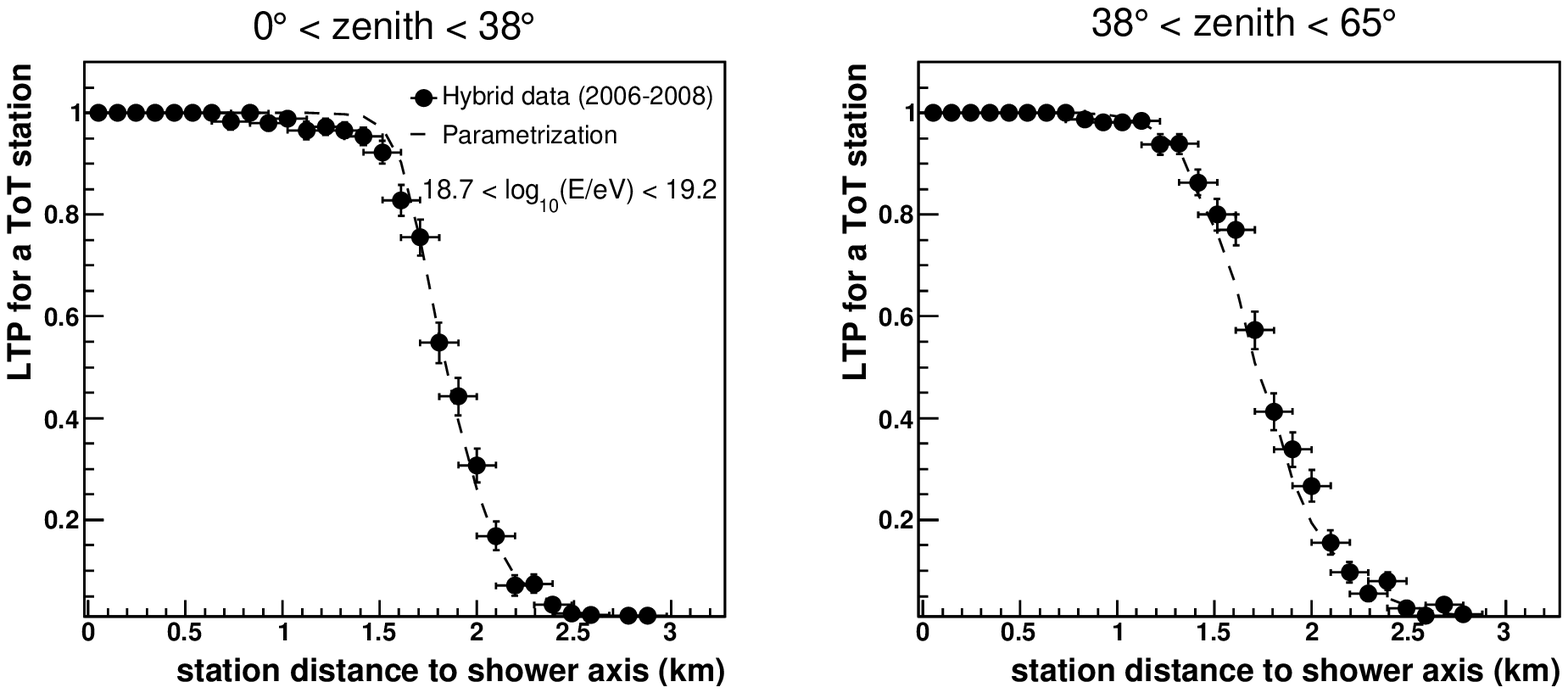}            
\caption{Comparison of simulation with hybrid data collected in two years. Zenith angles are split in two ranges 
0$^{\circ}$~-~38$^{\circ}$ (left) and 38$^{\circ}$~-~65$^{\circ}$ (right). 
From top to bottom the energy intervals are 
10$^{17.2}$~$<$~$E$~$<$~10$^{17.7}$~eV,   
10$^{17.7}$~$<$~$E$~$<$~10$^{18.2}$~eV, 
10$^{18.2}$~$<$~$E$~$<$~10$^{18.7}$~eV, 
10$^{18.7}$~$<$~$E$~$<$~10$^{19.2}$~eV.
}
\label{fig:sim-data-zen}
\end{center}
\end{figure}

For each energy interval considered, the agreement between data and 
simulation has also proven to hold in two zenith angle bands (0$^{\circ}$~-~38$^{\circ}$ and 38$^{\circ}$~-~65$^{\circ}$), see 
Fig.~\ref{fig:sim-data-zen}.

\subsection{Impact of weather effect on LTP}

\begin{figure}[p]
\begin{center}
    \includegraphics[width=0.99\textwidth]{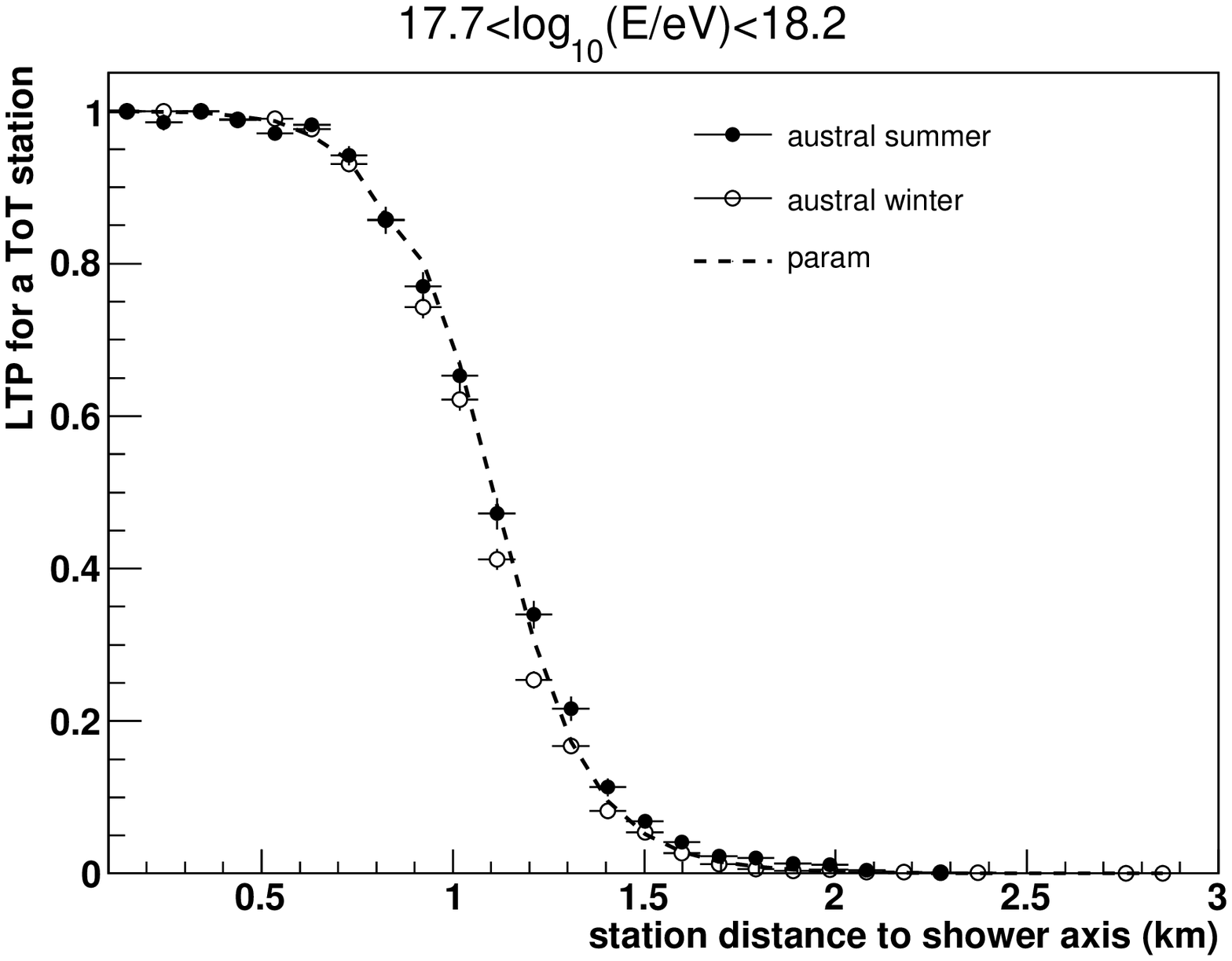}
    \includegraphics[width=0.99\textwidth]{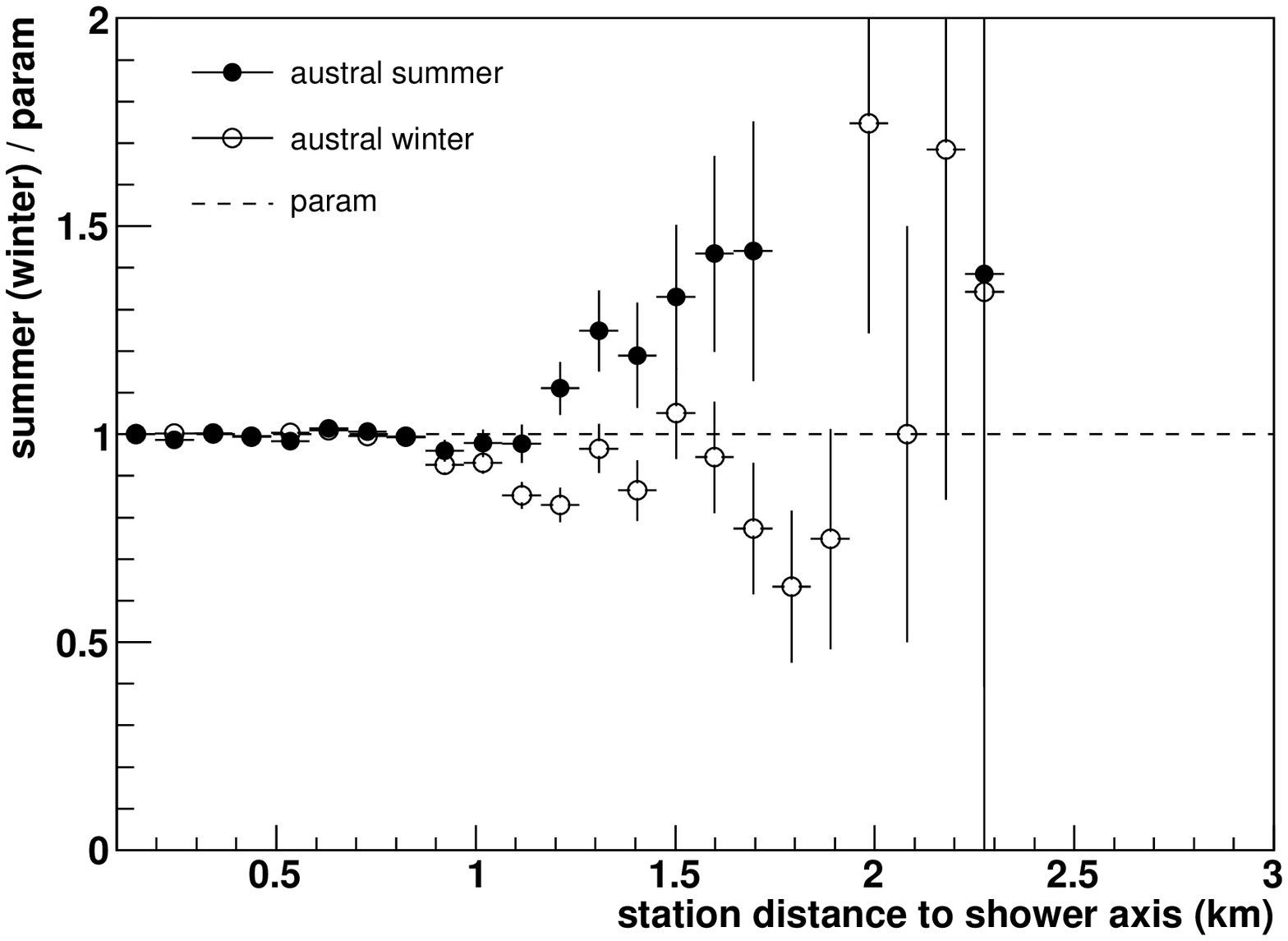}    
\caption{LTP functions from hybrid data at energy of about 10$^{18}$~eV 
for austral winter and austral summer compared to 
the parametrization derived in section~\ref{sec:simulation} (top) and ratio relative to the parametrization (bottom).}
\label{fig:weather}
\end{center}
\end{figure}

The effect of atmospheric variations (in pressure, temperature and air density) 
on extensive air showers development has been extensively studied with the  
surface detector data~\cite{weather}. 
A significant modulation
of the rate of events with the atmospheric variables, both
on a seasonal scale ($\sim$ 10\%) and on a shorter time scale ($\sim$ 2\% on
average during a day) has been observed.
This modulation is mainly explained as due
to the change with the air density
of the Moli\`ere radius near ground thus influencing the  
trigger probability and the rate of events above a fixed energy.
Hybrid data in the energy range around 10$^{18}$~eV have been used to investigate  
this effect on LTP. Data have been separated by  
season and are shown, together with the parametrization, for austral winter and austral summer,   
see Fig.~\ref{fig:weather}, top panel. The ratio of summer and winter relative to the parametrization 
is shown in the bottom panel. 
Results qualitatively match the expectation.  
Higher temperature at the ground, as for the austral summer, induces a reduction of 
the air density weakly enhancing the trigger probability at a given distance relative to 
all other seasons. 
Nevertheless the effect is almost negligible on the scale of the measurable trigger efficiency. 

\section{Summary and Conclusions}
\label{sec:conclusions}

In the previous sections we have introduced the concept of Lateral Trigger Probability function as a tool to characterize the single detector trigger efficiency. 
We have derived LTP functions for the particular case of the surface detector of the Pierre Auger Observatory  
using simulations.  We discussed their evolution with different physical parameters of air showers such as the energy, zenith angle and nature of the primary particle. 
We also investigated the impact of choosing different hadronic interaction models in the simulations.  
Furthermore, we estimated the LTP functions at different energies and zenith angles using hybrid data  
and showed that seasonal effects are visible in the trigger probabilities retrieved from data as expected from previous studies~\cite{weather}.

The good agreement between simulations and data over a wide energy range (between 10$^{17.5}$~eV and 10$^{19}$~eV) demonstrates the accuracy of the different aspects of the simulation procedure 
(i.e.\, air shower, detectors and trigger simulation) as well as the quality of the reconstruction 
obtained for hybrid data. These comparisons support and validate the use of simulated LTP functions in the estimate of the hybrid aperture described in~\cite{hybrid_aperture}.
Monitoring the LTP functions over a longer period of time can be used to study  
the long-term performance of the SD trigger for individual stations  
both above and below the acceptance saturation energy. 

As a final consideration, LTP functions can be derived at higher energies using SD-only data because, at energy above 
$\sim10^{18.5}~$eV, despite the statistics of hybrids becoming small,  
 the surface detector is fully efficient and the geometrical reconstruction is accurate. 
As mentioned in the Introduction, the probability of a high level trigger for the surface detector is a 
combination of single detector probabilities. Hence LTP functions provide a robust and simple method to estimate the energy or zenith angle 
dependence of SD acceptance for any arbitrary configuration.  This makes this technique a valuable tool to design other experiments and 
future enhancements of the Pierre Auger Observatory.

\section{Acknowledgements}
\label{sec:ack}

The successful installation and commissioning of the Pierre Auger Observatory
would not have been possible without the strong commitment and effort
from the technical and administrative staff in Malarg\"ue.

We are very grateful to the following agencies and organizations for financial support: 
Comisi\'on Nacional de Energ\'{\i}a At\'omica, 
Fundaci\'on Antorchas,
Gobierno De La Provincia de Mendoza, 
Municipalidad de Malarg\"ue,
NDM Holdings and Valle Las Le\~nas, in gratitude for their continuing
cooperation over land access, Argentina; 
the Australian Research Council;
Conselho Nacional de Desenvolvimento Cient\'{\i}fico e Tecnol\'ogico (CNPq),
Financiadora de Estudos e Projetos (FINEP),
Funda\c{c}\~ao de Amparo \`a Pesquisa do Estado de Rio de Janeiro (FAPERJ),
Funda\c{c}\~ao de Amparo \`a Pesquisa do Estado de S\~ao Paulo (FAPESP),
Minist\'erio de Ci\^{e}ncia e Tecnologia (MCT), Brazil;
AVCR, AV0Z10100502 and AV0Z10100522,
GAAV KJB300100801 and KJB100100904,
MSMT-CR LA08016, LC527, 1M06002, and MSM0021620859, Czech Republic;
Centre de Calcul IN2P3/CNRS, 
Centre National de la Recherche Scientifique (CNRS),
Conseil R\'egional Ile-de-France,
D\'epartement  Physique Nucl\'eaire et Corpusculaire (PNC-IN2P3/CNRS),
D\'epartement Sciences de l'Univers (SDU-INSU/CNRS), France;
Bundesministerium f\"ur Bildung und Forschung (BMBF),
Deutsche Forschungsgemeinschaft (DFG),
Finanzministerium Baden-W\"urttemberg,
Helmholtz-Gemeinschaft Deutscher Forschungszentren (HGF),
Ministerium f\"ur Innovation, Wissenschaft und Forschung, Nordrhein-Westfalen,
Ministerium f\"ur Wissenschaft, Forschung und Kunst, Baden-W\"urttemberg, Germany; 
Istituto Nazionale di Fisica Nucleare (INFN),
Istituto Nazionale di Astrofisica (INAF),
Ministero dell'Istruzione, dell'Universit\`a e della Ricerca (MIUR), 
Gran Sasso Center for Astroparticle Physics (CFA), Italy;
Consejo Nacional de Ciencia y Tecnolog\'{\i}a (CONACYT), Mexico;
Ministerie van Onderwijs, Cultuur en Wetenschap,
Nederlandse Organisatie voor Wetenschappelijk Onderzoek (NWO),
Stichting voor Fundamenteel Onderzoek der Materie (FOM), Netherlands;
Ministry of Science and Higher Education,
Grant Nos. 1 P03 D 014 30 and N N202 207238, Poland;
Funda\c{c}\~ao para a Ci\^{e}ncia e a Tecnologia, Portugal;
Ministry for Higher Education, Science, and Technology,
Slovenian Research Agency, Slovenia;
Comunidad de Madrid, 
Consejer\'{\i}a de Educaci\'on de la Comunidad de Castilla La Mancha, 
FEDER funds, 
Ministerio de Ciencia e Innovaci\'on and Consolider-Ingenio 2010 (CPAN),
Generalitat Valenciana, 
Junta de Andaluc\'{\i}a, 
Xunta de Galicia, Spain;
Science and Technology Facilities Council, United Kingdom;
Department of Energy, Contract Nos. DE-AC02-07CH11359, DE-FR02-04ER41300,
National Science Foundation, Grant No. 0969400,
The Grainger Foundation USA; 
NAFOSTED, Vietnam;
ALFA-EC / HELEN,
European Union 6th Framework Program,
Grant No. MEIF-CT-2005-025057, 
European Union 7th Framework Program, Grant No. PIEF-GA-2008-220240,
and UNESCO.



\bibliography{Bibliography}

\begin{thebibliography}{99}





\bibitem{engineering}
J. Abraham et al. [Pierre Auger Collaboration], Nuclear Instruments and Methods in  Physics Research A523 (2004) 50. 


\bibitem{FD_paper}
J. Abraham et al. [Pierre Auger Collaboration], Nuclear Instruments and Methods in Physics Research A620 (2010) 227.

\bibitem{hybrid_aperture}
P.Abreu et al. [Pierre Auger Collaboration],  Astroparticle Physics 34 (2011) 368. 


\bibitem{SD_acceptance}
J. Abraham et al. [Pierre Auger Collaboration], 
Nuclear Instruments and Methods in Physics Research A613 (2010) 29.


\bibitem{combined_spectrum} J. Abraham et al. [Pierre Auger Collaboration], Physics Letters B685 (2010) 239.
 
\bibitem{xmax}  J. Abraham  et al. [Pierre Auger Collaboration], Physical Review Letters 104 (2010) 091101.  


\bibitem{ltpicrc2005} D. Allard, 29th Int. Cosmic Ray Conf. (2005) arXiv:astro-ph/0511104v1.



\bibitem{corsika}
D. Heck {\em et al.} ,
{\it ``CORSIKA: A Monte Carlo Code to Simulate Extensive Air Showers''},
Report FZKA 6019, (1998).




\bibitem{thinning}
D. Heck and J. Knapp, Report FZKA 6097 (1998) Forschungszentrum Karlsruhe;
http://www-ik.fzk.de/heck/publications/


\bibitem{qgsjet}
S.Ostapchenko, Physics Letters B636 (2006) 40, 
 Physical Review D74 (2006) 014026. 

\bibitem{fluka}
A. Fass\`{o} et al. 
{\it FLUKA: a multi-particle transport code}, CERN-2005-10 (2005) INFN/TC\_05/11, SLAC-R-773;  
A. Fass\`{o} et al.  
{\it The physics models of FLUKA: status and recent developments}, 
Computing in High Energy and Nuclear Physics 2003 Conference (CHEP2003),  
La Jolla, CA, USA, March 24-28, 2003, (paper MOMT005) eConf C0303241 (2003) arXiv:hep-ph/0306267. 



\bibitem{geant4}
S. Agostinelli et al., 
Nuclear Instruments and Methods in  Physics Research A506 (2003) 250;  
IEEE Transactions on Nuclear Science  53 No. 1 (2006) 270.

\bibitem{billoir}
P. Billoir, Astroparticle Physics 30 (2008) 270. 


\bibitem{offline}
S. Argir\`{o} et al.,   
Nuclear Instruments and Methods in  Physics Research  A580 (2007) 1485.


\bibitem{dova_asym}
M.T. Dova, M.E. Manceñido, A.G. Mariazzi, H. Wahlberg, F. Arqueros D. García-Pinto 
Astroparticle Physics 31 (2009) 312.



\bibitem{sibyll}
Eun-Joo Ahn et al.,  Physical Review D80, (2009) 094003;
 R.S. Fletcher et al., Physical Review D50 (1994) 5710.




 





\bibitem{weather}
J. Abraham et al. [Pierre Auger Collaboration], Astroparticle  Physics 32 (2009) 89.
 
\end{thebibliography}


\newpage
\appendix
\section*{Appendix: LTP parametrization}

The LTP is fitted, as discussed in section~\ref{sec:simulation}, 
 to the following function:

\begin{equation}\tag{A.1}
\text{LTP(r)}  = \left\{ \begin{array}{ll}
\frac{1}{1+e^{-\frac{r-R_0}{\Delta R}}} & \textrm{$r\leq R_0$}\\ \\
\frac{1}{2}e^{C \cdot (r-R_0)} & \textrm{$r> R_0$}\\
\end{array} \right.
\end{equation}
with $R_{0}$ being the distance where the LTP is
equal to 0.5. \\ 
The dependences of fit parameters  $R_{0}$, $\Delta R$ and $C$ 
on energy and zenith angle can be parametrized by quadratic polynomials    
 in the variables $\cos\,\theta$ and $log_{10}(E/eV)$.
 The corresponding coefficients are given for 
proton, iron and photon primaries ($0^{\circ} < \theta < 65^{\circ}$), separately.
Concerning the accuracy of the parameters, 
 a change at the level of  $(1 \div 5)\%$ propagates approximately linearly in  
the returned value of the parametrization.

\subsection*{Proton showers}
The overall parametrization for proton primaries ($0^{\circ} < \theta < 65^{\circ}$)
is summarized in the following matrix equation:  
\begin{equation*}\label{protonTab}
\begin{split}
\frac{R_0}{\text{\text{km}}} =&
\left(\begin{array}{c}
       1  \\
       \cos\,\theta \\
       \cos^2\,\theta \\
       \end{array} \right)^{T}
       \cdot
    \left[          
\left( \begin{array}{rrr}
 4.30\cdot10^{1} & -6.21\cdot10^{0} &  2.09\cdot10^{-1} \\
-9.89\cdot10^{0} &  3.22\cdot10^{0} & -1.34\cdot10^{-1} \\
 -8.24\cdot10^{0} & -2.29\cdot10^{-1} &  3.11\cdot10^{-2} \\
       \end{array} \right)
\cdot \left( \begin{array}{c}
1 \\
\log_{10}(E/eV) \\
\log_{10}^2(E/eV) \\
            \end{array} \right) \right]
\\
\frac{\Delta R}{\text{km}} =&
\left(\begin{array}{c}
       1  \\
       \cos\,\theta \\
       \cos^2\,\theta \\
       \end{array} \right)^{T}
       \cdot
    \left[          
\left( \begin{array}{rrr}
-3.90\cdot10^{0} &  4.38\cdot10^{-1} & -1.15\cdot10^{-2} \\
1.19\cdot10^{1} & -1.37\cdot10^{0} &  3.82\cdot10^{-2} \\
-6.19\cdot10^{0} &  7.14\cdot10^{-1} & -1.99\cdot10^{-2} \\
\end{array} \right)
\cdot \left( \begin{array}{c}
1 \\
\log_{10}(E/eV) \\
\log_{10}^2(E/eV) \\
\end{array} \right) \right]
\\
\frac{\,C}{\text{km$^{-1}$}} =&
\left(\begin{array}{c}
       1  \\
       \cos\,\theta \\
       \cos^2\,\theta \\
       \end{array} \right)^{T}
       \cdot
    \left[          
\left( \begin{array}{rrr}
-3.28\cdot10^{2} & 3.48\cdot10^{1} & -9.16\cdot10^{-1} \\
-4.37\cdot10^{1}  & 3.96 \cdot10^{0}& -1.10\cdot10^{-1}  \\
 0~~~~& 0~~~~ & 0~~~~ \\
\end{array} \right) 
\cdot \left( \begin{array}{c}
1\\
\log_{10}(E/eV)\\
\log_{10}^2(E/eV)\\
\end{array} \right) \right]
\end{split}
\end{equation*}

\subsection*{Iron showers}
The overall parametrization for iron  primaries ($0^{\circ} < \theta < 65^{\circ}$)
is summarized in the following matrix equation: 
\begin{equation*}\label{IronTab}
\begin{split}
\frac{R_0}{\text{\text{km}}} =&
\left(\begin{array}{c}
       1  \\
       \cos\,\theta \\
       \cos^2\,\theta \\
       \end{array} \right)^{T}
       \cdot
    \left[          
\left( \begin{array}{rrr}
 4.90\cdot10^1 & -6.97\cdot10^0 & 2.33 \cdot10^{-1}\\
-9.23\cdot10^3 & 3.07\cdot10^0  & -1.30\cdot10^{-1} \\
-24.4\cdot10^3 & 1.69\cdot10^0 & -2.43\cdot10^{-2} \\
\end{array} \right)
\cdot \left( \begin{array}{c}
1 \\
\log_{10}(E/eV) \\
\log_{10}^2(E/eV) \\
\end{array} \right) \right]
\\
\frac{\Delta R}{\text{km}} =&
\left(\begin{array}{c}
       1  \\
       \cos\,\theta \\
       \cos^2\,\theta \\
       \end{array} \right)^{T}
       \cdot
    \left[          
\left( \begin{array}{rrr}
 -9.52\cdot10^{-1}& 6.81\cdot10^{-2} & 0~~~~  \\
 1.46\cdot10^{0}  & -1.04\cdot10^{-1}  & 0~~~~ \\
-9.32\cdot10^{-1}  & 6.36\cdot10^{-2} & 0~~~~\\
\end{array} \right)
\cdot \left( \begin{array}{c}
1 \\
\log_{10}(E/eV) \\
\log_{10}^2(E/eV) \\
\end{array} \right) \right]
\\
\frac{\,C}{\text{km$^{-1}$}} =&
\left(\begin{array}{c}
       1  \\
       \cos\,\theta \\
       \cos^2\,\theta \\
       \end{array} \right)^{T}
       \cdot
    \left[          
\left( \begin{array}{rrr}
 -8.82\cdot10^{2}& 9.50\cdot10^{1} & -2.56\cdot10^{0} \\
 3.83\cdot10^{2} & -4.40\cdot10^{1} & 1.24\cdot10^{0} \\
 0~~~~ & 0~~~~ & 0~~~~\\
\end{array} \right)
\cdot \left( \begin{array}{c}
1 \\
\log_{10}(E/eV) \\
\log_{10}^2(E/eV) \\
\end{array} \right) \right]
\end{split}
\end{equation*}

\subsection*{Photon showers}
The overall parametrization for photon  primaries ($0^{\circ} < \theta < 65^{\circ}$)
is summarized in the following matrix equation: 
 \begin{equation*}\label{photonTab}
\begin{split}
\frac{R_0}{\text{\text{km}}} =&
\left(\begin{array}{c}
       1  \\
       \cos\,\theta \\
       \cos^2\,\theta \\
       \end{array} \right)^{T}
       \cdot
    \left[          
\left( \begin{array}{rrr}
1.07\cdot10^{2}  & -1.31\cdot10^{1} & 3.89\cdot10^{-1} \\
 -2.46\cdot10^{2} & 2.90\cdot10^{1} & -8.30\cdot10^{-1}  \\
 1.47\cdot10^{2} & -1.70\cdot10^{1} & 4.78\cdot10^{-1} \\
       \end{array} \right)
\cdot \left( \begin{array}{c}
1 \\
\log_{10}(E/eV) \\
\log_{10}^2(E/eV) \\
            \end{array} \right) \right]
\\
\frac{\Delta R}{\text{km}} =&
\left(\begin{array}{c}
       1  \\
       \cos\,\theta \\
       \cos^2\,\theta \\
       \end{array} \right)^{T}
       \cdot
    \left[          
\left( \begin{array}{rrr}
9.03\cdot10^{0} &  -1.02\cdot10^{0} &  3.05\cdot10^{-2}\\
-2.76\cdot10^{1} & 3.15\cdot10^{0}  & -9.26\cdot10^{-2} \\
 2.46\cdot10^{1} & -2.82\cdot10^{0} & 8.25\cdot10^{-2}\\
\end{array} \right)
\cdot \left( \begin{array}{c}
1 \\
\log_{10}(E/eV) \\
\log_{10}^2(E/eV) \\
\end{array} \right) \right]
\\
\frac{\,C}{\text{km$^{-1}$}} =&
\left(\begin{array}{c}
       1  \\
       \cos\,\theta \\
       \cos^2\,\theta \\
       \end{array} \right)^{T}
       \cdot
    \left[          
\left( \begin{array}{rrr}
 -9.34\cdot10^{3} & 1.04\cdot10^{3} & -2.91\cdot10^{1}  \\
 2.60\cdot10^{4} & -2.91\cdot10^{3} &  8.10\cdot10^{1} \\
  -1.67\cdot10^{4} & 1.86\cdot10^{3} &  -5.17\cdot10^{1}\\
\end{array} \right) 
\cdot \left( \begin{array}{c}
1\\
\log_{10}(E/eV)\\
\log_{10}^2(E/eV)\\
\end{array} \right) \right]
\end{split}
\end{equation*}

\end{document}